\documentclass[12pt]{article}
\pdfoutput=1  

\usepackage{graphicx}
\usepackage{epsfig}
\usepackage{epstopdf}
\usepackage{amsmath}
\usepackage{amsfonts}
\usepackage{amssymb}
\usepackage{ccaption}
\usepackage[usenames]{color}

\usepackage[
      colorlinks=true,
      linkcolor=blue,  
      citecolor=blue,
      urlcolor=blue,    
      filecolor=blue,     
      linkcolor=blue,
      pdfstartview=FitV,
      pdftitle={},
        pdfauthor={Hubeny},
        pdfsubject={bulk probes},
        pdfkeywords={AdS/CFT},
        pdfpagemode=None,
        bookmarksopen=true    
      ]{hyperref}

\definecolor{labelkey}{gray}{0.1}
  \captionnamefont{\bfseries}
  \captiontitlefont{\small\sffamily}
  \captiondelim{: }
  \hangcaption

\def\sec#1{Section \ref{#1}}
\def\fig#1{Fig.\,\ref{#1}}
\def\req#1{(\ref{#1})}

\def\eps{\epsilon}

\def\ph{\varphi}
\def\p{\partial}

\def\iffs{\qquad \Longleftrightarrow \qquad}

\def\ands{\qquad {\rm and} \qquad}

\def\CA{{\cal A}}
\def\CB{{\cal B}}

\def\CH{{\cal H}}

\def\CL{{\cal L}}
\def\CM{{\cal M}}
\def\CN{{\cal N}}
\def\CO{{\cal O}}

\def\CR{{\cal R}}
\def\CS{{\cal S}}

\def\ZZ{\mathbb{Z}}

\def\RR{\mathbb{R}}

\def\l{\ell}
\def\p{\partial}
\def\Veff{V_{\rm eff}}
\def\rmin{r_{\ast}}
\def\zmax{z_{\ast}}
\def\rh{r_+}
\def\zh{z_+}
\def\sfcf{q}
\def\zd{{\dot z}}
\def\zdd{{\ddot z}}
\def\zdp{{\dot {\bar z}}}
\def\zp{{\bar z}}
\def\zpp{{\bar {\bar z}}}

\def\rhd{{\dot \rho}}
\def\rhdd{{\ddot \rho}}
\def\rhdp{{\dot {\bar \rho}}}
\def\rhp{{\bar \rho}}
\def\rhpp{{\bar {\bar \rho}}}

\def\znFG{{\tilde z}}  
\def\znFGh{{\tilde z}_+}
\def\znFGmax{{\tilde z}_{\ast}}

\def\expval#1{{\langle \, #1  \, \rangle}}
\def\corr#1#2{{\langle \CO( #1 ) \, \CO( #2 )  \rangle}}

\numberwithin{equation}{section}

\title{{\bf \Large Extremal surfaces as bulk probes in AdS/CFT}}

\author{\normalsize 
Veronika E. Hubeny\footnote{veronika.hubeny@durham.ac.uk} \\
\small \sl   Centre for Particle Theory \& Department of
Mathematical Sciences,
\\[-1.5mm]
\small \sl Science Laboratories, South Road, Durham DH1 3LE, United Kingdom. \\
}

\begin{document}

\setlength{\baselineskip}{16pt}
\begin{titlepage}
\maketitle
\begin{picture}(0,0)(0,0)
\put(350,253){DCTP-12/15}
\end{picture}
\vspace{-26pt}

\begin{abstract}

Motivated by the need for further insight into the emergence of AdS bulk spacetime from CFT degrees of freedom, we  explore the behaviour of probes represented by specific geometric quantities in the bulk.
We focus on geodesics and $n$-dimensional extremal surfaces in a general static asymptotically AdS spacetime with spherical and planar symmetry, respectively.
While our arguments  do not rely on the details of the metric,  we illustrate some of our findings explicitly in spacetimes of particular interest (specifically AdS, Schwarzschild-AdS and extreme Reissner-Nordstrom-AdS).

In case of geodesics, we find that for a fixed spatial distance between the geodesic endpoints, spacelike geodesics at constant time can reach deepest into the bulk.  
We also present a simple argument for why, in the presence of a black hole, geodesics cannot probe past the horizon whilst anchored on the  AdS boundary at both ends.
The reach of an extremal $n$-dimensional surface anchored on a given region depends on its dimensionality, the shape and size of the bounding region, as well as the bulk metric.  We argue that for a fixed extent or volume of the boundary region, spherical regions give rise to the deepest reach of the corresponding extremal surface.  Moreover, for physically sensible spacetimes, at fixed extent of the boundary region, higher-dimensional surfaces reach deeper into the bulk.
Finally, we show that in a static black hole spacetime, no extremal surface (of any dimensionality, anchored on any region in the boundary) can ever penetrate the horizon.

\end{abstract}
\thispagestyle{empty}
\setcounter{page}{0}
\end{titlepage}

\renewcommand{\thefootnote}{\arabic{footnote}}


\tableofcontents

\section{Introduction}
\label{intro}

The gauge/gravity duality\footnote{
For definiteness we'll consider the prototypical case of the AdS/CFT correspondence \cite{Maldacena:1997re} 
which relates the four-dimensional $\CN=4$ Super Yang-Mills  (SYM) gauge theory to a IIB string theory (or supergravity) on asymptotically AdS$_5 \times S^5$ spacetime.
} has been famously fruitful at revealing universal features of strongly coupled field theories via the dual gravitational description.   This has led to very successful programs such as AdS/QCD and  AdS/CMT  which even bear links to present-day experiments.
On the other hand, the promise which the gauge/gravity correspondence has yet to fulfill is the potentiality of answering long-standing questions of quantum gravity by recasting them into non-gravitational, field-theoretic language.  One prerequisite to such reformulation is a thorough understanding of the gauge/gravity map:  how do the gravitational degrees of freedom emerge from the field theory?  In particular how do the field theoretic degrees of freedom organize themselves so as to give rise to bulk locality, at least at the classical level?

In order to address such questions, it is useful to understand which features of the CFT are most sensitive to the bulk geometry.  For example, given a specific bulk location, what quantities in the CFT should we examine in order to learn about the physics at the specified location?  
Where on the background spacetime of the CFT are these  `probes' localized?
As a starting point, we want to learn how much of the bulk can given CFT observables `see'.   
 More specifically, suppose we consider some finite region $\CB$ of the boundary CFT where we have full knowledge of the relevant CFT ``data", namely we know $\expval{\CO(x)}$, $\corr{x}{y}$, etc., for all $x,y \in \CB$, at least for some chosen set of field theory operators $\CO$.  The question we wish to address is: what is the maximal {\it bulk} region $\CM_{\CB}$ for which the bulk geometry is uniquely specified by the CFT data in $\CB$?
Conversely, suppose we wish to determine the bulk metric fully in some region $\CM$.  What is the minimal set of CFT data (in particular minimal boundary region on which we need to  specify such data) $\CB_{\CM}$ that will allow complete specification of the metric in $\CM$?
These are not easy questions, since bulk locality is not manifest -- nor indeed well-understood -- in the dual CFT.\footnote{
Early investigations of bulk locality from various perspectives include
\cite{Banks:1998dd,Polchinski:1999ry,Susskind:1998vk,Giddings:1999jq,Horowitz:2000fm,Hamilton:2005ju}
whereas more recent developments and reviews are given in e.g.\
\cite{Gary:2009mi,Heemskerk:2009pn,Kabat:2011rz,Fitzpatrick:2011jn,Heemskerk:2012mn}.
}
  On the other hand, answering them would lend us insight into how holography encodes bulk locality, and in turn into the emergence of spacetime.

In exploring these questions, we wish to be minimalist in our assumptions about the bulk spacetime.  In particular, will not assume any field equations for the bulk, analyticity of the metric, etc.  The reason for this is the following. 
One might optimistically suppose that if we can extract enough data about the  bulk geometry at some spacelike slice of the bulk to provide an initial condition for bulk evolution (and if we can liberally assume that we know the boundary conditions at all times), then we could declare that we automatically know the entire bulk geometry simply by evolving.  In this we are effectively condensing the description of the full spacetime to that of the initial condition, which thoroughly obscures the intricacies of emergence of bulk time.
More drastically, if the metric were analytic and we knew its exact form in any open neighborhood, then we would automatically know it everywhere.  But such tricks miss the point of decoding the boundary-to-bulk map.  Not only do they require the knowledge of a piece of spacetime to infinite precision which is not practically possible, but more importantly they don't offer much intuition about how the gauge/gravity map works. 
To penetrate closer to the core of the AdS/CFT mechanism it is more useful to restrict ourselves to using only local information about the bulk geometry and try to understand which probes see the bulk metric most directly.

Why do we elevate the bulk metric, as opposed to other bulk attributes, to be the primary quantity of interest?  There are several answers to this question.
From a pragmatic standpoint, it is the minimal set of dynamical variables which is guaranteed to exist (in more than 3 bulk dimensions):  The bulk action always contains the Einstein-Hilbert term (along with a negative cosmological constant), and in fact this sector in isolation constitutes a consistent truncation of the full theory.  
Relatedly, if we wish to extract the configuration of any other matter fields, we may do so in addition to, but not instead of, the bulk metric, since the very notion of bulk regions where these matter fields reside requires knowing the bulk geometry.
Finally, if there is any sort of realistic matter, as opposed to just test probes, its backreaction on the geometry will guarantee that its presence will bear its signatures in the metric -- so in that sense by extracting the details of the bulk geometry with sufficient precision, we are detecting much of the physics taking place in the bulk.

That said, our study deals with probes of the geometry, not physical quantities which backreact on it.  From the bulk standpoint, particularly natural constructs are various geometrical quantities such as geodesics or higher-dimensional extremal surfaces, since their position and geometrical attributes do not require any specification of coordinates, spacetime foliation, etc. -- they are defined in a fully covariant manner.
The basic idea for using such constructs to uncover the bulk geometry is simply the following.  For given set of boundary conditions, the location (and in turn the length/area/volume, etc.) of these quantities is determined by the geometry.  This means that they encode some aspect of the bulk geometry along their support.  By comparing these quantities for nearby boundary conditions we can isolate the bulk locations which contribute most significantly to the difference in their attributes.  As indicated below, in some circumstances one can invert this relation to extract the bulk metric explicitly.
  Fortuitously, such  geometric constructs find their use in the dual field theory, and in fact many correspond to very natural and fundamental CFT probes.  Here we mention a few notable examples (for a review, see e.g.\ \cite{Hubeny:2010ry}).
\begin{itemize}
\item {\it Geodesics:} Correlators of high-dimension operators can be expressed in the WKB approximation heuristically as 
\begin{equation}
\corr{x}{y} \sim e^{-m \, \CL(x,y)}
\label{}
\end{equation}	
where $m$ is the mass of the bulk field corresponding to $\CO$ and $\CL(x,y)$ is a regularized proper length along a spacelike geodesic connecting $x$ and $y$.  Since such geodesic passes through the bulk, this correlator is sensitive to the corresponding  region of the bulk geometry.
This was used to identify the CFT signature of the black hole singularity 
\cite{Fidkowski:2003nf,Festuccia:2005pi}, building on  prior work of \cite{Balasubramanian:1999zv,Louko:2000tp,Kraus:2002iv}, by
 considering the insertion points $x$ and $y$  on two disconnected boundaries corresponding to the two asymptotic regions of an eternal black hole.\footnote{
In fact as later demonstrated in \cite{Balasubramanian:2007qv}, by examining the detailed analytic structure of such correlation functions, one can tell apart spacetimes with and without horizons, and even distinguish such subtle differences as the fuzzball picture of the black hole \cite{Mathur:2005zp,Skenderis:2008qn} and a genuine eternal black hole geometry.
Indeed, 
one can use similar correlators to discern even more subtle signals of the bulk geometry from behind the black hole horizon, such as explored in \cite{Freivogel:2005qh} to investigate aspects of inflationary cosmology within AdS/CFT. 
}
However, the impressive amount of information extractible from these correlators comes with a cost:  since the insertion points are not located on the same boundary, accessing this information as genuine signal in the CFT requires analytic continuation.
That in turn imposes unwanted restriction on the spacetime.

This shortcoming is avoided in the framework of \cite{Hubeny:2006yu}, which considers the structure of singularities of generic Lorentzian correlators, observing that these correlators  exhibit light-cone singularities when the operator insertion points are connected by a  null geodesic.  This geodesic can lie along the boundary (which gives rise to the familiar light cone singularity in the field theory), or it can traverse the bulk, emerging within\footnote{
As proved by \cite{Gao:2000ga}, in a certain wide class of spacetimes with timelike conformal boundary, any Òfastest null geodesicÓ connecting two points on the boundary must lie entirely within the boundary. 
The vacuum state of the CFT plays a distinguished role in this context, as all null geodesics through pure AdS bulk take the same time to reach the antipodal point as the boundary null geodesics.
} the boundary light cone; the latter give rise to the {\it bulk-cone singularities}.
The structure of these singularities was recently examined by \cite{Erdmenger:2011aa} (following \cite{Amado:2008hw,Erdmenger:2011jb}) in context of moving mirror in AdS.
As pointed out in \cite{Hubeny:2006yu}, bulk-cone singularities can be used to extract part of the spacetime geometry, even including the location of the bulk event horizon formation in a collapse geometry.   This was subsequently used in \cite{Hammersley:2006cp, Bilson:2008ab} for recovering the metric within a simple class of static spherically symmetric spacetimes, by numerical and analytical means respectively.

\item {\it Extremal surfaces:} 
The expectation value of a Wilson-Maldecena loop
is related to the area of two-dimensional extremal surface describing a string world-sheet ending on the corresponding contour in the boundary of AdS \cite{Maldacena:1998im,Rey:1998ik}.
Similarly, the entanglement entropy \cite{Nishioka:2009un},\cite{Casini:2011kv} associated to a given region $\CA$ in the boundary is related to the area of a co-dimension 2 extremal\footnote{
As argued in \cite{Hubeny:2007xt}, such extremal surface is in fact related to  light-sheet constructions of the covariant entropy bound \cite{Bousso:1999xy} in the bulk spacetime.
Note that  in 2+1 bulk dimensions, the bulk duals of equal-time correlators, Wilson loops, and entanglement entropy probes all degenerate to constant-time spacelike geodesics.
} surface ending on $\p \CA$.  This was first developed by  \cite{Ryu:2006bv,Ryu:2006ef} for static configurations and later generalized to time-dependent situations by \cite{Hubeny:2007xt}, which emphasized its use in extracting bulk geometries and demonstrated how this proposal may be used to understand the time evolution of entanglement entropy in a time varying QFT state dual to a collapsing black hole background.
The analysis of thermalization (serving as a toy model of ``quantum quench") initiated in \cite{Hubeny:2007xt} was recently extended by 
\cite{AbajoArrastia:2010yt,Aparicio:2011zy} in 3-dimensional bulk, in
\cite{Albash:2010mv} for 4 dimensions, and 
\cite{Balasubramanian:2010ce,Balasubramanian:2011ur} in 3,4 and 5 dimensions,
the latter having used entanglement entropy as well as equal time correlators and Wilson loop expectation values.
(See also \cite{Albash:2011nq,Albash:2012pd} for other recent explorations of holographic entanglement entropy as a probe in different contexts.)
\end{itemize}

One might naively wonder why we don't simply use the expectation value of the boundary stress tensor to extract the bulk metric, since after all it is the boundary stress tensor which couples to the bulk metric.  Indeed, several previous approaches utilized this substantially.  In the holographic renormalization group scheme of \cite{deHaro:2000xn}, one can reconstruct the asymptotic behaviour of the bulk geometry using Fefferman-Graham expansion around the boundary. This expansion however is not guaranteed to converge deep in the bulk, and in fact would generically lead to singular geometries.
In the fluid/gravity correspondence \cite{Bhattacharyya:2008jc} the problem of singularities is avoided by working in the long-wavelength regime, where the boundary stress tensor prescribes the bulk geometry down to well inside the bulk event horizon.  This may then appear attractively good probe of the geometry, but it comes at the cost of disallowing any sharp variations in the bulk geometry; in a sense, we can `probe' so deep into the bulk mainly because there's not much happening in the transverse directions, so the main variation (in the radial direction) is very similar to the exact solution of a static black hole we know already.  
So neither of these methods would be able to discern a localized deformation in the geometry, say, somewhere in the vicinity of the event horizon.

A more pedestrian way to state the shortcoming of the stress tensor expectation value is that it only knows about the asymptotic fall-off of the bulk metric.  In fact, it is trivial to construct examples with matter which have the same boundary stress tensor but drastically different bulk metric.  One such example is a spherically symmetric distribution of matter.  The asymptotic metric fall-off only knows about the total ADM mass, whereas the bulk geometry is of course sensitive to the radial density profile and time dependence of this matter.
This shortcoming is of course avoided if we don't restrict ourselves merely to the expectation values of the stress tensor, but are allowed to consider its higher point functions as well.  
However, once we allow this level of complexity, we might as well consider $n$-point functions of  more general operators without worrying about the tensor structure, as well as more non-local observables.

Therefore we will focus on CFT quantities such as correlators, Wilson loop expectation values, and entanglement entropy, which provide examples of our probes, i.e.\ ``data" that we'll assume we have access to in some region of the CFT.  Rather than furthering the program of using these to extract the bulk geometry, we  wish explore the more general question of {\it which of these is the best-suited} for extracting the bulk geometry.
As mentioned above,
for asking matter-of-principle questions such as what is the best CFT probe of bulk geometry, it is convenient to separate this issue from 
the field equations.  We will therefore refrain from imposing Einstein's equations.  Instead, we consider any physically sensible geometry (which would be supported by some physical matter), and ask what is the most economical way to decode parts of this geometry from the boundary data.
On the other hand, in order to make progress, we will focus on static and spherically or translationally symmetric geometries.  The expectation is that these cases, albeit highly symmetric, are sufficiently indicative of the general case, in terms of elucidating which CFT probe is the best-suited to probing the bulk geometry.

Of special interest in our considerations will be causally nontrivial spacetimes corresponding to a black hole in AdS.
It is clear by causality that no timelike or null curve in the bulk can probe inside the black hole horizon while being anchored to the  AdS boundary at both ends.  No such causal obstruction applies to spacelike curves of course, however the story is more interesting for spacelike geodesics or higher-dimensional extremal surfaces.  We will first show that in fact in static spherically symmetric spacetimes spacelike geodesics likewise cannot penetrate the horizon, and subsequently generalize this statement to a much larger family of higher-dimensional extremal surfaces anchored on the boundary.
As we remark below, this is no longer true in general for time-evolving spacetimes; indeed it is easy to find counter-examples, such as given in \cite{Hubeny:2002dg}.

The structure of this paper is as follows.  In \sec{s:geods} we focus on geodesic probes of static, spherically symmetric, asymptotically AdS spacetimes.  Apart from the result mentioned above that these cannot penetrate black holes, we argue that the best-suited geodesics for probing deepest into the bulk are the spacelike ones localized at constant time.  Motivated by this observation, we go on to consider extremal surfaces anchored on the boundary at constant time.\footnote{
These are often referred to as ``minimal surfaces" in the literature; however since we are working in Lorentzian spacetime where the area of a spacelike surface can be decreased by deforming it in the timelike direction, we will maintain the terminology ``extremal".}  For simplicity, we also switch from global AdS to Poincare AdS, namely consider extremal surfaces in general static asymptotically AdS spacetimes which are translationally invariant in the boundary directions.
In \sec{s:ExtSurf} we examine these surfaces anchored on variously-shaped boundary regions of various dimensionalities, exploring how these attributes affect 
the depth to which 
such surfaces can reach in the bulk.  We also consider the effect that spacetime deformations have on this reach.
We conclude by presenting an argument that out of all possible shapes with fixed area or extent, spherical regions allow for the greatest reach.
\sec{s:ExtSurfHor} focuses on spacetimes with horizons, demonstrating that extremal surfaces of any dimensionality, and anchored on arbitrarily shaped region in the boundary, cannot penetrate event horizon in our general class of spacetimes.  Further discussion and summary of the main results appears in \sec{s:Discussion}.  
A preview of some of these results were given in \cite{Hubeny:2010ry}, and some related results appeared in e.g.\ \cite{Hubeny:2006yu,Ryu:2006ef,Hubeny:2007xt}; however we have attempted to keep the presentation  self-contained and pedagogical.

\section{Geodesic probes}
\label{s:geods}

In this section we consider CFT probes related to bulk geodesics anchored on the AdS boundary at both endpoints.  We will refer to such geodesics as {\it probe} geodesics, since they can be more directly associated to a natural CFT probe.  Our main concern is how sensitive are these probe geodesics to the  geometry of a given bulk spacetime.
For simplicity, we will restrict our considerations to asymptotically (globally) AdS, static, spherically symmetric bulk spacetimes.  By a suitable choice of coordinates we can write the line element as
\begin{equation}
ds^2= - f(r) \, dt^2 + h(r) \, dr^2 + r^2 \, d\Omega^2
\label{metgen}
\end{equation}	
where $f$ and $h$ are a-priori arbitrary\footnote{
Ultimately, their functional dependence will be constrained by the Einstein's equations and boundary conditions; but as motivated in the Introduction, we will consider arbitrary spacetimes, which are not necessarily solutions to any specified field equations.
} radial functions with the large-$r$ behaviour determined by the AdS asymptotics, namely
$f(r) \sim \frac{1}{h(r)} \sim r^2 +1$ as ${r\to \infty}$, 
 and we are for now keeping the spacetime dimension $d+1$ arbitrary.  Note that in the above expression we have fixed the AdS radius to unity, tantamount to measuring all distances in AdS units.

We want to ask the following questions:
\begin{itemize}
\item {\it How much of the bulk is accessible to the set of all probe geodesics? } For spherically symmetric spacetimes \req{metgen}, this will be characterized by the minimal radius $\rmin$ to which probe geodesics can reach.  For pure AdS and small deformations thereof, the answer is of course the full spacetime so $\rmin=0$; however, there exist physically relevant spacetimes, such as ones with a black hole, where some part of the bulk cannot be reached by geodesics with both endpoints anchored on the (same) boundary.
\item In the latter set of cases where some part of the bulk remains inaccessible, {\it which probe geodesics can reach the deepest? } Since all geodesics in \req{metgen}  can be characterized by energy $E$ and angular momentum $L$, as well as the discrete parameter $\kappa$ distinguishing null,  spacelike, and timelike ones, this question boils down to finding the optimal $E$, $L$, and $\kappa$ which minimize the corresponding $\rmin$.
\item A further refined question is, {\it for some restricted region on the boundary from which we allow the probe geodesics to emanate, what part of the bulk is accessible?}
\item It is also of interest to know {\it what is the (regularized) proper length of a given probe geodesic?}  We expect that the larger this quantity is, the harder it would be to extract  from the corresponding CFT probe.  On the other hand, shorter-length geodesics typically reach less deep into the bulk, so they carry less information about the  geometry.  
\end{itemize}

To preview the main results, we will see that in any spacetime of the form \req{metgen}:
\begin{enumerate}
\item
For given $E$ and $L$, spacelike geodesics necessarily probe deeper than null ones.
\item 
For restricted angular separation of the endpoints, the spacelike geodesic which reaches the deepest is the  one moving in constant-time slice, with smallest allowed angular momentum consistent with the restriction on the endpoints.
\item
When the geometry contains a black hole, probe geodesics cannot reach past the horizon.
However, by suitable adjusting of the parameters we can find a probe geodesic which reaches arbitrarily close to the horizon.  
\item
Nevertheless, if we impose an upper bound on the angular separation of the endpoints 
or the regularized proper distance 
of such a probe geodesic, the restricted probe geodesic only reaches to some finite distance from the horizon.  
In practice, this is not too drastic a constraint, and the near-horizon region remains accessible for sensible restrictions.
\item
Null probe geodesics, on the other hand, can only penetrate to the unstable circular orbit around the black hole, which is typically some ${\mathcal O}(1)$ distance from the horizon.
\end{enumerate}

The remainder of this section consists of a pedagogical derivation of these results.
In \sec{s:gensetup} we describe the geodesic in terms of 1-D motion in an effective potential for arbitrary spacetime of the form \req{metgen}, focusing on the minimal radius $\rmin$ reached by such a geodesic, its endpoints, and the regularized proper length.
To make our discussion more specific, we then separate the set of possibilities into two classes:  globally static spacetimes (which are causally trivial), addressed in \sec{s:globstat}, and spacetimes with an event horizon, addressed in \sec{s:statBH}, where we give the general proof that probe geodesics cannot reach past the horizon.  For illustrative purposes, we include the particular example of pure AdS in \sec{s:globstat} and three black hole spacetimes (BTZ, Schwarzschild-AdS$_5$, and extremal Reissner-Nordstrom-AdS$_5$) in  \sec{s:statBH}. 
 
\subsection{General set-up}
\label{s:gensetup}

Let us consider a general geodesic in a static spherically symmetric spacetime \req{metgen}.  Using spherical symmetry, we can fix angular coordinates such that a given geodesic remains confined to the equator of the $S^{d-1}$.  Denoting the azimuthal angle $\ph$, we can then write the tangent vector along this geodesic as
\begin{equation}
p^a = \dot{t} \, \p_t^a + \dot{r} \, \p_r^a + \dot{\ph} \, \p_{\ph}^a 
\label{geodgen}
\end{equation}	
where $\dot{} \equiv \frac{d}{d\lambda}$ with $\lambda$ denoting the affine parameter along the geodesic.
Using the usual constants of motion induced by the Killing fields $\p_t^a$ and $\p_{\ph}^a$, namely the energy $E \equiv - p_a \, \p_t^a = f(r) \, \dot{t}$ and angular momentum $L \equiv p_a \, \p_{\ph}^a = r^2 \, \dot{\ph}$, and the norm of the tangent vector $\kappa \equiv p_a \, p^a$ (fixed to $\kappa = +1, 0, -1 $ for spacelike, null, and timelike geodesics, respectively), we can recast the radial geodesic equation in terms of a 1-d motion in an effective potential $\Veff$, 
\begin{equation}
\dot{r}^2 + \Veff(r) = 0 \ , \qquad \qquad
\Veff(r) = \frac{1}{h(r)} \, \left[ -\kappa - \frac{E^2}{f(r)} + \frac{L^2}{r^2} \right] \ .
\label{Veffgen}
\end{equation}	
Since classically defined geodesic motion requires $\dot{r}^2 \ge 0$,
the condition for such a geodesic to reach the boundary is then simply $\Veff(r\to \infty) <0$, and the deepest into the bulk (i.e.\ smallest $r$) that such a geodesic reaches, denoted $\rmin$, is given by the largest root of $\Veff(r)$.  The main goal of this section is to study the behaviour of $\rmin$.

Using the AdS asymptotics, we can immediately see that at large $r$, for non-null geodesics $\Veff$ is dominated by $\Veff \sim -\kappa\, r^2$, so timelike geodesics ($\kappa = -1$) cannot reach the boundary for any finite value of $E$ and $L$. 
On the other hand, all spacelike geodesics (in this asymptotic region) necessarily reach the boundary for any value of $E$ and $L$.  For null geodesics, since $\kappa =0$, by constant rescaling the affine parameter we can fix $E =1$ and parametrize the geodesic by a single parameter $\l \equiv L/E$, so that at large $r$, $\Veff \sim \l^2 - 1$.  This means that in order for null geodesics to reach the boundary,\footnote{
Depending on the spacetime, there may or may not exist null geodesics with $\l^2 >1$: in pure AdS, it is easy to see that $\l^2 >1$ null geodesics simply do not exist, since 
$\Veff(r) > \frac{\l^2-1}{r^2} > 0$ for $\l^2 >1$.  On the other hand, in a black hole geometry, null geodesics can exist for arbitrary $\l$, but for $\l^2 >1$ they cannot reach the boundary.  In fact, for very large $\l^2$, they cannot emerge far from the vicinity of the horizon:
$r_{\rm max}-\rh \sim 1/\l^2$.
} we need to take $\l^2 < 1$ (the limiting value of $\l=\pm 1$ would correspond to the null geodesic staying on the boundary, i.e.\ $\rmin = \infty$).  Finally, note that the high-energy limit of a spacelike geodesic, $E \to \infty$ with $\l = L/E$ fixed, corresponds to the null geodesic with angular momentum $\l$ and infinitely rescaled affine parameter.\footnote{
Note that one has to be careful about order of limits, as illustrated by the following example: consider spacelike geodesic with say $L = 2E$ and take the limit $E \to \infty$ with $L/E$ fixed.  This is a null geodesic with $\l=2$ which therefore cannot reach the boundary, yet all spacelike geodesics in this family were anchored on the boundary.
(However, $\rmin \to \infty$ as $E \to \infty$.)}
We will therefore only consider spacelike geodesics ($\kappa = 1$ with arbitrary $E$ and $L$) and null geodesics ($\kappa = 0$ with $E^2 < L^2$ or equivalently $\l^2 <1$).  In order to capture the behaviour of both spacelike and null geodesics, we will keep describing the null ones in terms of the redundant notation of $E$ and $L$, and simply set $\kappa = 0$ in the resulting expressions.

\paragraph{Reach of spacelike versus null geodesics:}
We first give a very simple argument why certain spacelike geodesics can reach deeper into the bulk (i.e.\ have smaller $\rmin$) than given null geodesics.
The argument uses the assumption that $h(r)$ and $f(r)$ are positive in the region of interest, $\rmin \le r < \infty$, whose proof we postpone till \sec{s:statBH}.
Consider any null geodesic, specified by $\l$.  This is described by an effective potential \req{Veffgen} with $\kappa = 0$ and $L = \l \, E$ for any $E>0$. Denote its largest root by $\rmin^{(\kappa=0)}$.  Now, a spacelike geodesic with the same $E$ and $L$ will have its effective potential lowered everywhere by $\frac{1}{h(r)} > 0$.  This means that in particular at $\rmin^{(\kappa=0)}$, the effective potential for this spacelike geodesic will still be negative, and therefore its largest root $\rmin^{(\kappa=1)}$ will have to occur at smaller value of $r$, i.e.\ $\rmin^{(\kappa=1)} < \rmin^{(\kappa=0)}$.  

We emphasize that this argument holds for general spacetimes, as long as $h(r)>0$ for $r \ge \rmin^{(\kappa=0)}$.  This is automatically satisfied when the spacetime is globally static, so that $h(r)>0$ for all $r \ge 0$.  As discussed in \sec{s:globstat}, in such a case, both sets of geodesics can probe to arbitrarily small radii $\rmin$ simply by choosing the angular momentum to be small enough.  In a causally non-trivial case, $h(r)$ may be negative, but we will argue \sec{s:statBH} that for any probe geodesic   $\rmin$ is bounded from below in such a way that the $h(r)<0$ region can never be reached.

It is also easy to see from \req{Veffgen} the effects of shifting $E$ and $L$ whilst keeping the other parameters fixed: 
\begin{itemize}
\item If we increase $L^2$   while keeping $E$ fixed, $\Veff$ increases at all $r$ where $h(r)>0$, so $\rmin$ likewise increases.  This is to be expected since there is greater centrifugal barrier.
\item On the other hand, as $E^2$ increases with $L$ fixed, $\Veff$ decreases at all $r$, so $\rmin$ decreases.  
\item However, as $E^2$ increases with $\ell=L/E$ fixed, $\Veff$ increases in the vicinity of the turning point, so $\rmin$ increases.  
\end{itemize}

We can see the last statement more clearly by writing \req{Veffgen} for the spacelike 
case as
\begin{equation}
\Veff^{(\kappa=1)}(r) 
	= \frac{1}{h(r)} \, 
	\left(   - 1 + E^2 \, \left[ - \frac{1}{f(r)} + \frac{\ell^2}{r^2} \right]\right) \ ,
\label{}
\end{equation}	
so that $\rmin^{(\kappa=1)}$ occurs at a value of $r$ for which
$\left[  - \frac{1}{f(r)} + \frac{\ell^2}{r^2} \right] = \frac{1}{E^2}$,
which is positive.  This means that the coefficient of $E^2$ in $\Veff^{(\kappa=1)}(r) $ is some function of $r$, which is positive at $r=\rmin$.  Hence if we lower $E^2$, the effective potential will decrease at the original turning point $r=\rmin$, which in particular means that the new\footnote{
Note that although the term in square brackets can take both positive and negative values depending on $r$, we can see that decreasing $E^2$ cannot suddenly introduce new larger roots of $\Veff$, essentially by running the previous argument backwards: raising $E^2$ increases the value of $\Veff$ near all its roots, but since $\Veff(r=\infty) < 0$, this operation would increase $\rmin$ rather than getting rid of it entirely.
} turning point will shift to lower value of $r$.  
This implies that to minimize $\rmin^{(\kappa=1)}$ at fixed $\ell$, we need to minimize $E^2$,  namely consider the case $E=0$.

We can in fact now figure out what the minimal radius is.    As long as $h(r)>0$ for all $r$, we can probe the full spacetime.  Indeed, as is clear from \req{Veffgen}, setting $E=0$ and $\kappa=1$, we immediately see that $\rmin = |L|$, which can be taken to $r=0$ when $L=0$.
As we will show in \sec{s:statBH}, if there is an event horizon at some $r= \rh$, then we can probe down to it.

\paragraph{Geodesic endpoints and proper length:}
Having addressed part of our first two questions, namely how much of the bulk is accessible to probe geodesics and which ones probe deepest, we now turn to the remaining two questions, which concern the endpoints and proper length of the geodesic.
In particular, suppose we have only a finite region in the CFT at our disposal, within which we can anchor our geodesic endpoints.  We would then be interested in finding the `optimal' probe geodesic within this restricted set of geodesics.  To that end, we need to specify the endpoints of a general probe geodesic.

Recall that we wish to focus on probe geodesics whose endpoints at $\lambda = \pm \infty$ are anchored on the boundary $r=\infty$.
Due to the time-translational and rotational symmetry of the spacetime \req{metgen}, the relevant data on the boundary are the temporal and angular distances between the two geodesic endpoints, 
\begin{equation}
\Delta t \equiv 2 \int_{\rmin}^{\infty} \frac{E}{f(r)} \, g(r) \, dr 
\qquad {\rm and} \qquad
\Delta \ph \equiv 2\int_{\rmin}^{\infty} \frac{L}{r^2} \, g(r) \, dr \ ,
\label{Dtphi}
\end{equation}	
where we have simplified the notation somewhat by defining
\begin{equation}
g(r) \equiv \sqrt{\frac{h(r)}{\kappa + \frac{E^2}{f(r)} - \frac{L^2}{r^2} }} = \frac{1}{\sqrt{-\Veff(r)}} = \frac{1}{|\dot{r}(r)|} \ .
\label{gdef}
\end{equation}	
Note that in each case both the integrand and the lower limit of integration $\rmin$ depend on the specific geodesic, characterized by the constants of motion $E$ and $L$, as well as on the spacetime, described by the functions $f(r)$ and $h(r)$.  
At large $r$ the integrands of \req{Dtphi} are proportional to $1/r^3$ for spacelike geodesics and $1/r^2$ for null geodesics, so that at the upper limit the integrals converge.  Moreover, it is easy to see that in the absence of unstable circular orbits, the integrals \req{Dtphi} also converge at the lower limit.  In particular, in the general case of $\Veff'(\rmin) \ne 0$, the we can write 
$\Veff(r) \approx \Veff'(\rmin) \, (r - \rmin)$, so that 
using $g(r)  = \frac{1}{\sqrt{-\Veff(r)}}$ we have the integrands near $r\approx \rmin$ scale with $\frac{1}{\sqrt{r-\rmin}}$.

For null geodesics the requirement that $\Veff(r)<0$ for all $r>\rmin$ implies that 
$\frac{L^2}{r^2}  < \frac{E^2}{f(r)} $, so by comparing the integrands in \req{Dtphi},
we can easily see that 
\begin{equation}
\Delta t \ge \ell \,  \Delta \ph \ .
\label{}
\end{equation}	
This is however already guaranteed by the stronger bound $\Delta t \ge \Delta \ph$ implied by a theorem of Gao \& Wald \cite{Gao:2000ga} which states that (subject to certain energy conditions) one can't propagate faster through the bulk than along the boundary.
Slightly stronger conditions can be obtained by considering the variation of the geodesic endpoints under varying the parameters.   
For example,  by varying the angular momentum $\l$ of null geodesics in any geometry, we obtain
\begin{equation}
\frac{\delta \Delta t}{\delta \l} = \l \,  \frac{\delta \Delta \ph}{\delta \l} \ , 
\label{}
\end{equation}	
and an analogous relation may be obtained for spacelike geodesics.

Finally, let us briefly turn to the proper length along a spacelike geodesic.  Since such a geodesic reaches the boundary, its proper length is by definition infinite; however we can easily regulate it.  
Simply imposing a large-radius cutoff at $r= R$, the proper length along the part of the geodesic with $r \le R$ is given by
\begin{equation}
{\cal L}_R = 2 \int_{\rmin}^{R} g(r) \, dr  \ .
\label{LpropR}
\end{equation}	
Following standard procedure, we then define the regularized proper length ${\cal L}_{\rm reg}$ by a background subtraction method as the large-$R$ limit of the difference between ${\cal L}_{\rm reg}$ in the spacetime of interest and ${\cal L}_{\rm reg}$ for the geodesic in pure AdS with the same endpoints.  In practice, this amounts to stripping off the $R$-dependent part from ${\cal L}_R$.

To make more definite statements about the behaviour of $\rmin$, $\Delta \ph$, $\Delta t$, and ${\cal L}_{\rm reg}$ for various spacetimes and parameters of a probe geodesic, let us now consider the case of spacetimes with and without horizons separately.  We start with the simpler case of causally trivial spacetimes in \sec{s:globstat}, and then turn to the more interesting black hole spacetimes in \sec{s:statBH}.

\subsection{Globally static spacetimes}
\label{s:globstat}

When the spacetime \req{metgen} is globally static, the Killing vector $\p_t^a$ is timelike everywhere, so $f(r) > 0$ for all $r \ge 0$, and Lorentzian signature then simultaneously forces $h(r) > 0$ for all $r \ge 0$.  
For convenience, we will consider the cases of radial geodesics with $L=0$ and non-radial geodesics with $L \ne 0$ separately.

\paragraph{$L = 0$ case:}
Let us first consider the special case of radial geodesics.  For $L=0$, the effective potential \req{Veffgen} is manifestly negative everywhere, so the geodesic continues all the way from the boundary to the origin.  The origin $r=0$ is a regular point, so the geodesic simply passes through, and continues back to $r=\infty$ on the antipodal point of the sphere.  Since the spacetime is globally static, it is causally trivial, so both geodesic endpoints lie on the same boundary.  The angular difference between the endpoints is simply $\Delta \ph = \pi$,
 while the time delay depends on the energy and the spacetime.  For null geodesics, the energy cancels out (as it must), and the time delay is simply given by
\begin{equation}
\Delta t_{\kappa=0, L=0} = 2 \int_0^\infty \sqrt{\frac{h(r)}{f(r)}} \, dr \ ,
\label{Dtnull}
\end{equation}	
whereas for spacelike geodesics \req{Dtphi} gives
\begin{equation}
\Delta t_{\kappa=1, L=0} = 2 \int_0^\infty \sqrt{\frac{h(r)}{f(r)}} \, \sqrt{\frac{1}{1+ \frac{f(r)}{E^2} }} \, dr \ .
\label{Dtspacelike}
\end{equation}	
Note that as $E \to \infty$ in the spacelike geodesic case, we recover the null geodesic result.  
It is now easy to compare $\Delta t$ between the spacelike and null case, for arbitrary globally static spacetimes.  In particular, since the integrand in \req{Dtspacelike} is always less than the integrand in \req{Dtnull}, we see that $\Delta t_{\kappa=1, L=0} < \Delta t_{\kappa=0, L=0}$, as we would expect.  Moreover, within the family of spacelike geodesics, $\Delta t_{\kappa=1, L=0}$ decreases with $E$; for $E=0$, $\Delta t_{\kappa=1, L=0, E=0} =0$, as expected.  This is intuitively obvious, since faster geodesics traveling the same spatial trajectory arrive in shorter time.

\paragraph{$L \ne 0$ case:}
We now turn to the more general case of $L \ne 0$.
Since $h(r) > 0$ for all $r \ge 0$, in looking for solution of $\Veff(\rmin) = 0$, we can ignore the overall $\frac{1}{h(r)}$ factor in \req{Veffgen}.  The minimal radius $\rmin$ can then be found by solving
\begin{equation}
-1 - \frac{E^2}{f(r)} + \frac{L^2}{r^2} = 0 \qquad {\rm at} \qquad r = \rmin \ .
\label{rmineq}
\end{equation}	
Since $L \ne 0$, there must always exist a solution $\rmin>0$ since at small $r$ the third term on the LHS of \req{rmineq} dominates, making the LHS large and positive, whereas at large $r$ the dominant first terms makes the LHS negative.  
It is easy to see that as we tune $L$ for given $E$, $\rmin$ ranges over all positive values.  In particular as $f$ varies between $f(r=0)= f_0$ and $f(r\to \infty) \sim r^2$, we have the small-$L$ and large-$L$ regime solutions:
\begin{equation}
\rmin \approx \frac{|L|}{\sqrt{1+\frac{E^2}{f_0}}} \  \ {\rm as} \ \ L \to 0
\qquad {\rm and} \qquad
\rmin \approx |L| \  \ {\rm as} \ \ L \to \infty \ .
\label{}
\end{equation}	

Unfortunately, in this case it is no longer as straightforward to compare $\Delta t$ and $\Delta \ph$ between null and spacelike geodesics, unlike in the $L=0$ case.  This is because both the integrand, and the lower limit of integration, in \req{Dtphi} is larger for a null geodesic case than  for the spacelike geodesic with same $E$ and $L$, so the overall integral depends more sensitively on the details.

\paragraph{Explicit expressions for pure AdS:}
To get better intuition, let us consider the simplest case of pure AdS, where the relevant expressions can be obtained in closed form.  We have $f(r) = h(r)^{-1} = r^2 +1$, for which case the effective potential for null and spacelike geodesics respectively simplifies to 
\begin{equation}
\Veff^{(\kappa = 0)}(r) = L^2 - E^2 + \frac{L^2}{r^2}
	\ands
\Veff^{(\kappa = 1)}(r) = -r^2 + (L^2 - E^2 - 1) + \frac{L^2}{r^2}	\ .
\label{VeffAdS}
\end{equation}	
We can easily find the minimal radius reached for any given geodesic.  For null geodesics,
\begin{equation}
\rmin^{(\kappa = 0)} = \sqrt{\frac{L^2}{E^2 - L^2}}
= \frac{|\l |}{\sqrt{1-\l^2}} \ ,
\label{rminAdSnull}
\end{equation}	
whereas for spacelike geodesics,
\begin{equation}
\rmin^{(\kappa = 1)} = \sqrt{ \frac{1}{2} \, \left[-( E^2 - L^2+1) + \sqrt{(E^2 - L^2+1)^2 +4L^2} \right] } 	\ .
\label{rminAdSsl}
\end{equation}	
For a spacelike geodesic with $E=0$, as in any globally static spacetime, the minimal radius must be at 
\begin{equation}
\rmin^{(\kappa = 1, E=0)} = |L| \ ,
\label{rminAdSslE0}
\end{equation}	
which  is indeed easy to verify directly from \req{rminAdSsl}.
In fact, the full radial trajectory of such $E=0$ geodesic, with $\rmin$ occurring at $\ph=0$ can be written in a very simple closed-form expression as
\begin{equation}
r^2(\ph) = \frac{L^2}{\cos^2 \ph - L^2 \, \sin^2 \ph} \ .
\label{}
\end{equation}	
From the form of \req{rminAdSsl}, we can easily confirm that at fixed $L$, increasing $E$ has the effect of lowering $\rmin^{(\kappa=1)}$.
On the other hand, at fixed $\l=L/E$, increasing $E$ increases $\rmin^{(\kappa=1)}$.

Let us now examine the relation between $\rmin$ and $(\Delta t, \Delta \ph)$.
By direct integration of \req{Dtphi}, 
we find that for a spacelike geodesic with arbitrary $E$ and $L$, 
\begin{equation}
\Delta \ph 
= \frac{\pi}{2}+ \sin^{-1} \left[ \frac{E^2 -L^2+1}{\sqrt{(E^2 -L^2+1)^2 + 4L^2 }} \right]
\label{}
\end{equation}	
and  
\begin{equation}
\Delta t = \frac{\pi}{2}+ \sin^{-1} \left[ \frac{E^2 -L^2-1}{\sqrt{(E^2 -L^2-1)^2 + 4E^2 }} \right]	\ .
\label{}
\end{equation}	
As an aside, we note that
$ E \, \cot \Delta t - L \, \cot \Delta \ph= 1$  for $E,L>0$.
In the special case of $E=0$ spacelike geodesics, we find $\Delta t = 0$ and $\Delta \ph$ is related to the angular momentum as
\begin{equation}
\Delta \ph = \frac{\pi}{2}+\sin^{-1}\frac{1-L^2}{1+L^2}
	= 2 \, \cot^{-1} |L| \ ,
\label{}
\end{equation}	
whereas for null geodesics, we have $\Delta t = \Delta \ph = \pi$ independently of $\l$.

We can also easily confirm that for fixed $\Delta \ph$ (leaving $\Delta t$ arbitrary), $\rmin$ is minimized at $E=0$ and grows monotonically with $E^2$.  This implies that if one has access to CFT data only in a certain region $\Delta \ph$, we can probe more of the bulk geometry with spacelike geodesics than with null geodesics.

For future reference, the regularized proper length along  a spacelike geodesic, given by \req{LpropR} with the universal divergent piece $\ln(4 R^2)$ stripped off, is 
\begin{equation}
{\cal L}_{\rm reg} = - \ln \sqrt{(E^2 -L^2+1)^2 + 4L^2} \ .
\label{}
\end{equation}	
For $E=0$ geodesics, this simplifies to ${\cal L}_{\rm reg} = - \ln \left(1+L^2\right) $, which vanishes for radial geodesics.
For fixed $L$, we see that the geodesic which minimizes $\rmin$, namely the constant-time one with $E=0$, in fact maximizes its length ${\cal L}_{\rm reg} $.  The same conclusion holds for fixed $\Delta \ph $.

\subsection{Spacetimes with an event horizon}
\label{s:statBH}

We now consider asymptotically-AdS spacetimes with a spherical black hole, i.e.\ spacetimes which are spherically symmetric and static, but not globally static.  The event horizon is a Killing horizon where the Killing field $\p_t^a$ has zero norm.  At this radial position, which we'll denote by $\rh$, $f(\rh)=0$.  
Although the static coordinates as written in \req{metgen} are singular on the horizon, we can easily pass to regular (e.g.\ ingoing Eddington) coordinates by defining
\begin{equation}
v = t + r_{\ast} \ \ , \ \ {\rm where } \qquad dr_{\ast} = \sqrt{\frac{h(r)}{f(r)}} \, dr
\label{}
\end{equation}	
upon which the line element \req{metgen} becomes
\begin{equation}
ds^2 = -f(r) \, dv^2 + 2 \, \sqrt{f(r) \, h(r) } \, dv \, dr + r^2 \, d\Omega^2
\label{}
\end{equation}	
and the conserved energy along a geodesic is $E = f(r) \, \dot{v} - \sqrt{f(r) \, h(r) } \, \dot{r}$.
The effective potential for geodesics of course still retains the same form as in \req{Veffgen}, namely
\begin{equation}
\Veff(r) = \frac{1}{h(r)} \, \left[ -\kappa - \frac{E^2}{f(r)} + \frac{L^2}{r^2} \right] \ .
\end{equation}	

Let us now consider the general properties of $\Veff(r)$, in particular the position of $\rmin$, which depends both on the spacetime specification $f(r)$ and $h(r)$, and on the geodesic (i.e.\ $\kappa$, $E$, and $L$).  Our present strategy will be the following: given a generic spacetime of the kind described above, we wish to explore what kinds of geodesics are possible on this spacetime.  In particular, what is the minimum $\rmin$ achievable by any probe geodesic (i.e.\ one with both endpoints anchored to the boundary), and for what $E,L,\kappa$ is it realized?  

On the horizon $r=\rh$ where $f(\rh)=0$, we need $f(\rh) \, h(\rh)  > 0$ to keep the metric well-defined and Lorentzian.  This means that the only term which survives in the effective potential at that point is 
\begin{equation}
\Veff(\rh) = - \frac{E^2}{f(\rh) \, h(\rh) } < 0 \ \ \ \  \forall  \ \ E \ne 0 \ .
\label{Veffrh}
\end{equation}	
Let us for now assume that $E>0$ (this simultaneously covers the cases of $E<0$ by flipping the time, and we'll return to the special case of $E=0$ later).
Then there are two qualitatively distinct possibilities for the behaviour of $\Veff$:
\begin{enumerate}
\item
$\Veff(r) < 0$ for all $r>\rh$.  Then the geodesic crosses the horizon and either there is a turning point inside the horizon at $\rmin < \rh$, or the geodesic continues to $r=0$.  In case of singularity at $r=0$, the geodesic ends there; otherwise $\rmin=0$ and the geodesic merely passes through a smooth origin.
\item
$\Veff(r) \ge 0$ for some $r>\rh$.  Then there are two\footnote{
In the special case when $\Veff(r)$ is nowhere positive, the zeros at  $\Veff =0$ degenerate.
} zeros of the effective potential outside the horizon, the larger of which corresponds to the turning point $\rmin > \rh$.  Such a geodesic never enters the black hole.
\end{enumerate}
Let us consider each of these cases in turn.

\begin{figure}
\begin{center}
\includegraphics[width=1.1in]{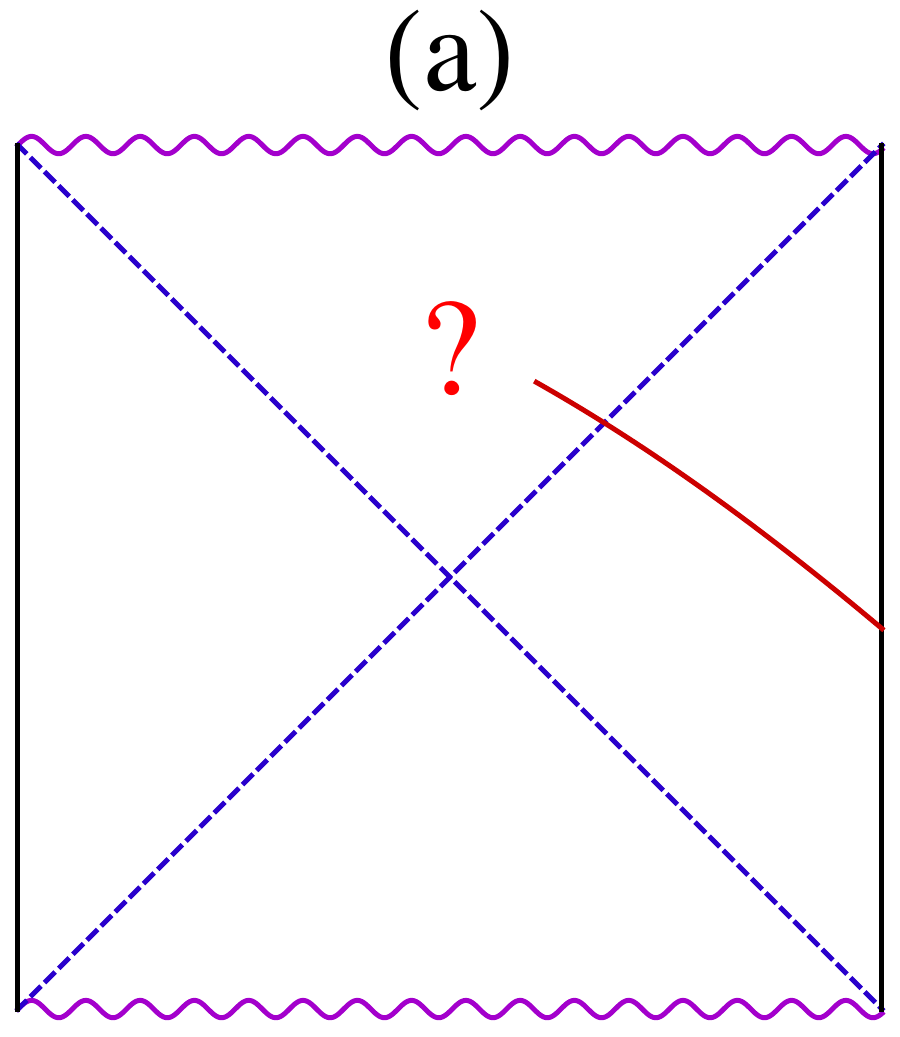}
\hspace{0.2cm}
\includegraphics[width=1.1in]{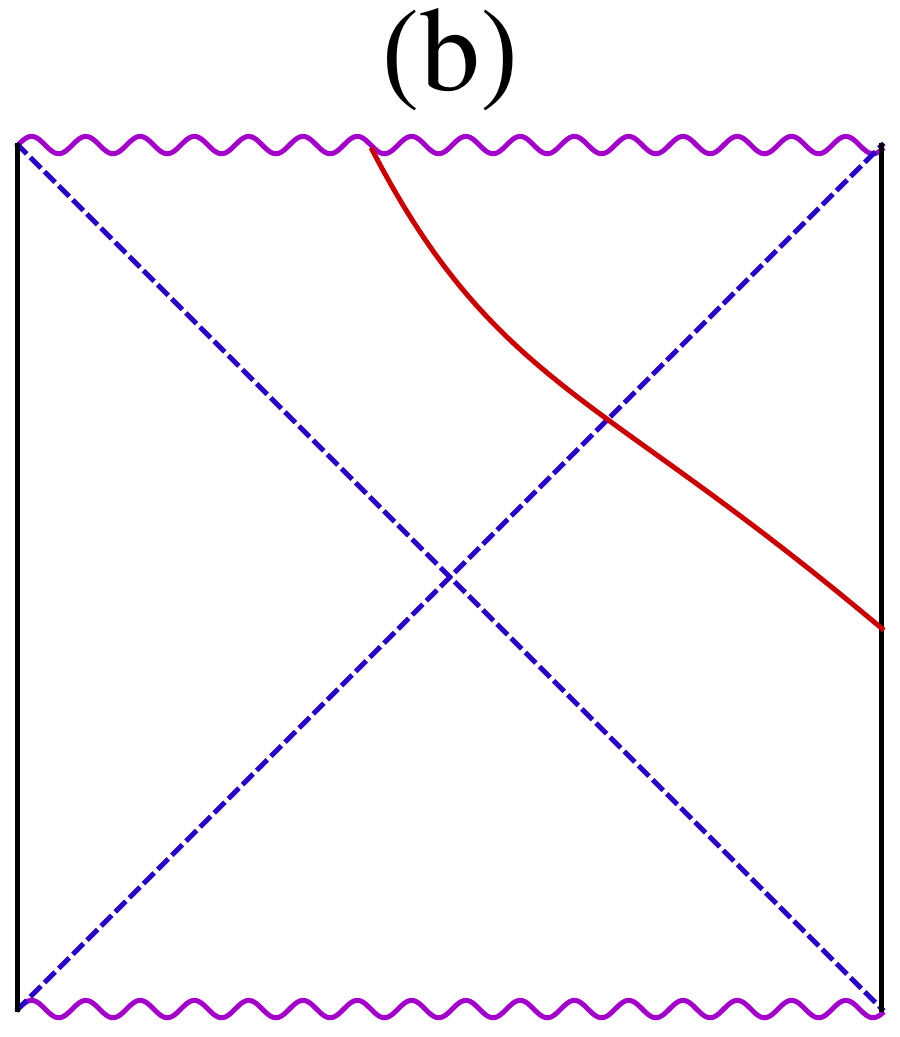}
\hspace{0.2cm}
\includegraphics[width=1.1in]{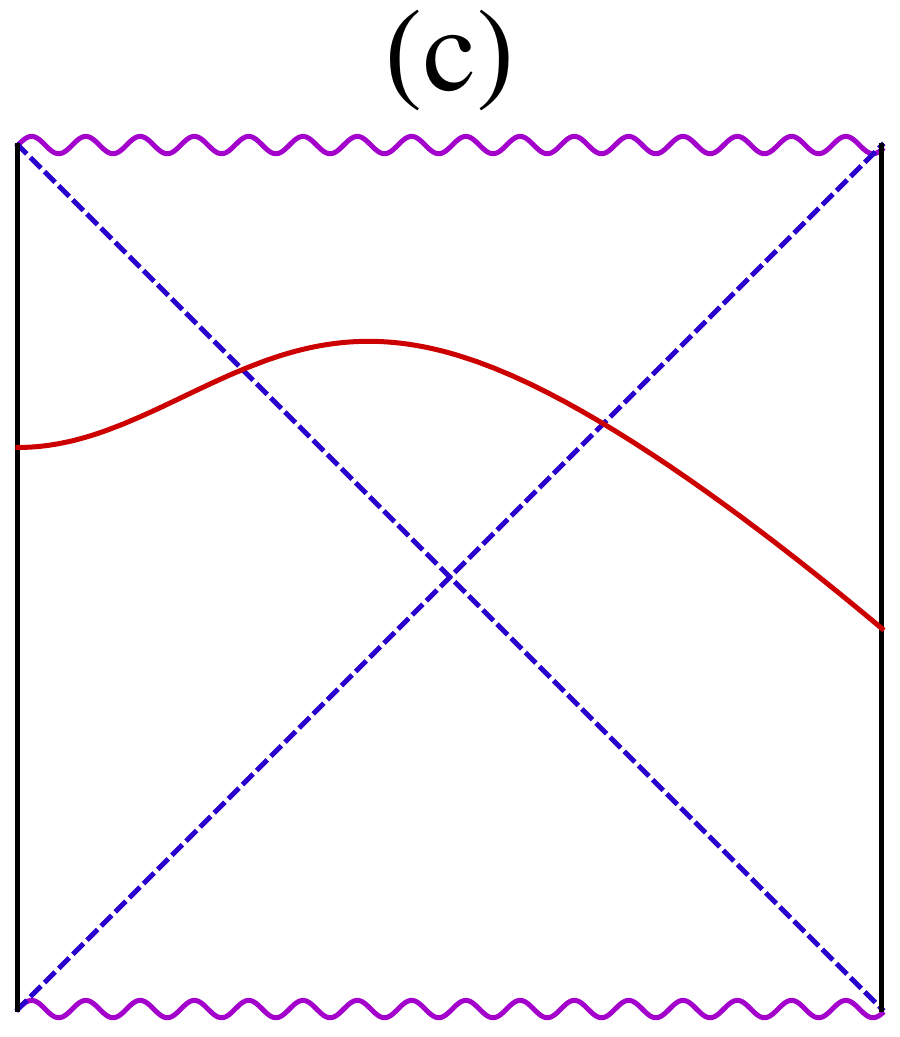}
\hspace{0.2cm}
\includegraphics[width=1.1in]{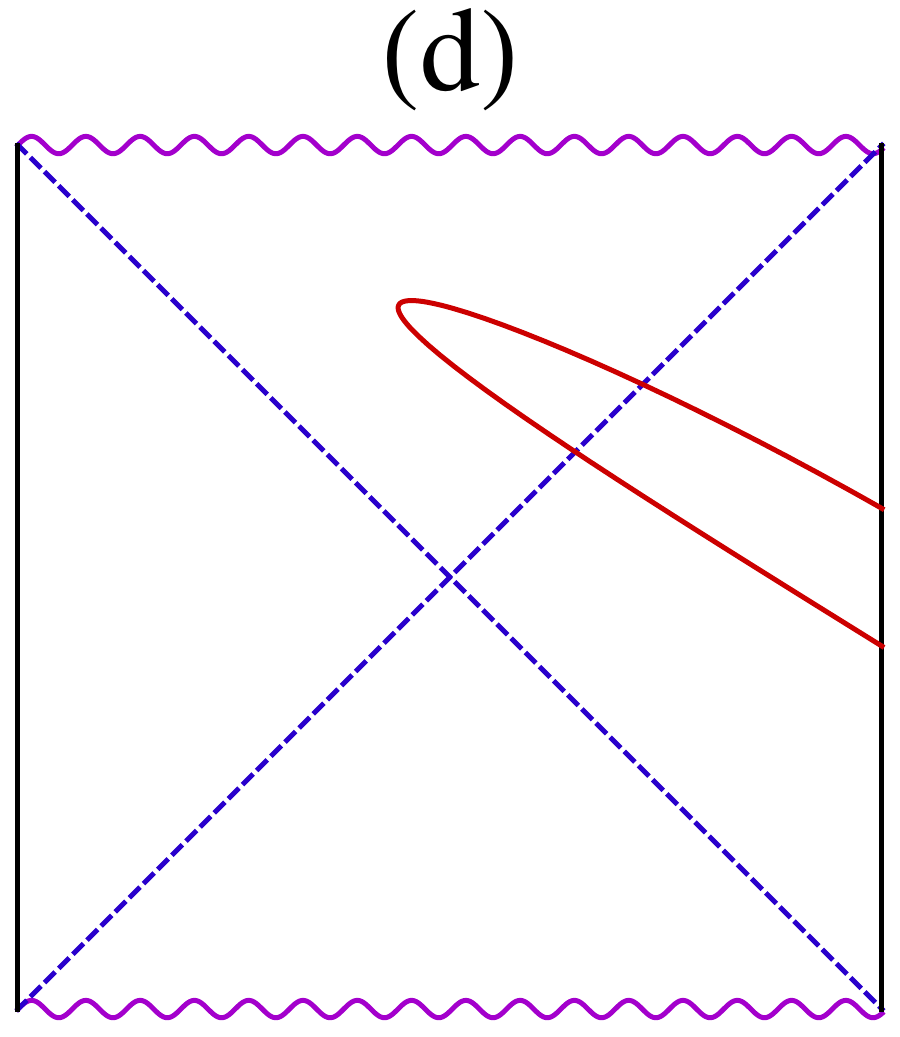}
\hspace{0.2cm}
\includegraphics[width=1.1in]{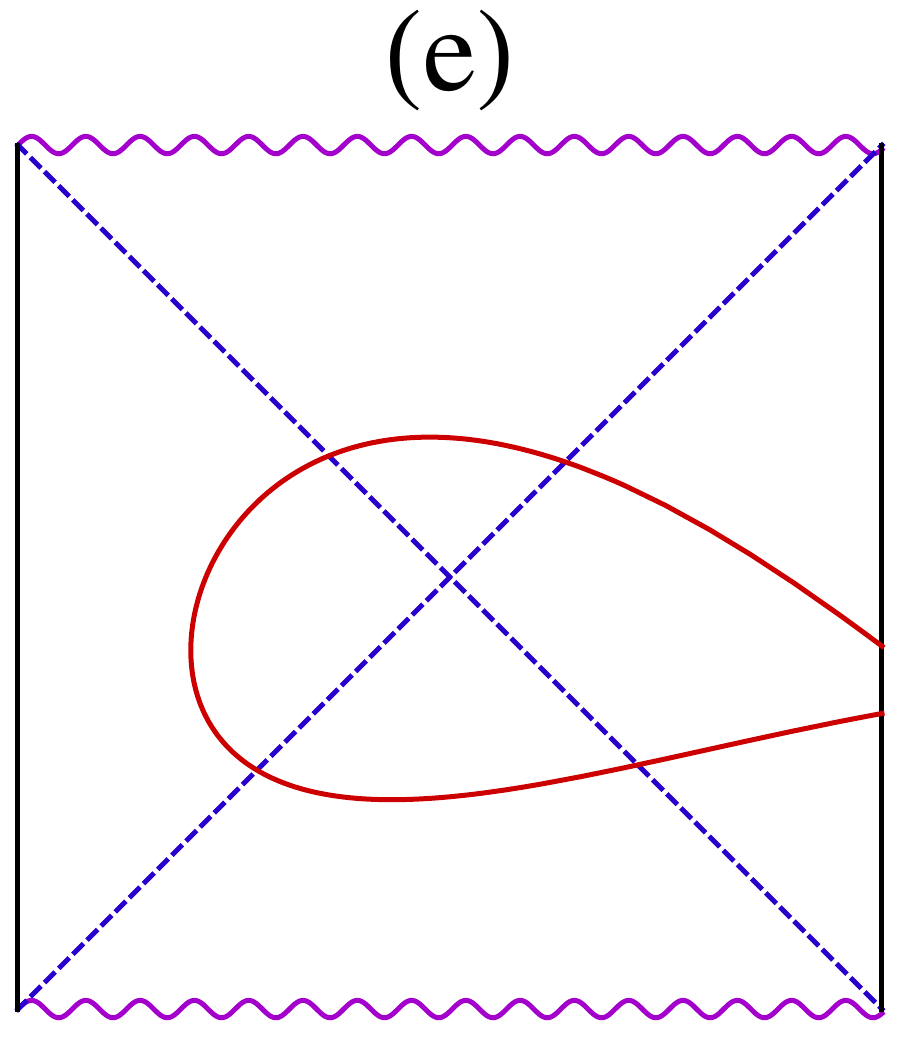}
\caption{
Sketch of possible behaviour of spacelike geodesics (red curves) on a Penrose diagram of an AdS black hole.$^{12}$ \,
Suppose a spacelike geodesic has crossed the future horizon (a).  There are several qualitatively different a-priori possibilities: it can end at the singularity (b), it can continue to the boundary of the other asymptotic region (c), it can return to the same boundary through the future horizon (d), or it can return to the same boundary via the past horizon (e).  We argue in the text that only (b) and (c) are viable possibilities.
}
\label{f:SAdSPD}
\end{center}
\end{figure}

\paragraph{Case 1 ($0 \le \rmin < \rh$):}
We will show that although such a geodesic enters the black hole, it can never return to the {\it same} asymptotic region.  Of course, this is obvious for timelike and null geodesics because of causal constraints, but here we will see that it is true even for spacelike geodesics which a-priori had no such constraints. 

Naively, there are several qualitatively distinct possibilities, illustrated in \fig{f:SAdSPD}, for what can happen to a geodesic which enters a black hole,\footnote{
For definiteness we consider a Schwarzschild-AdS-like causal structure with a spacelike curvature singularity, but this is not essential to our arguments, which rely only on the region around the horizon.
} i.e.\ crosses the future event horizon as indicated in \fig{f:SAdSPD}(a).  It can fall into the curvature singularity (b) or continue on to another asymptotic region (c).  Both of these cases are admissible, but neither corresponds to a {\it probe} geodesic, since the other endpoint  is not anchored on the same boundary.  

To restrict attention to potential probe geodesics, we are led to consider cases such as (d) or (e) where both endpoints are pinned at the same boundary; but as we now argue, these are inconsistent with the requirements of a geodesic. 
In particular, since in our spacetime $\p_v^a$ 
is a Killing field, $E$ is a conserved quantity along {\it any} geodesic.
Let us consider this constant of motion $E$ at the future horizon: 
\begin{equation}
E 
 =
\left[ f(r) \, \dot{v} - \sqrt{f(r) \, h(r) } \, \dot{r} \right]_{\rh} 
= - \sqrt{f(\rh) \, h(\rh)} \ \dot{r} > 0
\label{Eonrh}
\end{equation}	
where the second equality holds since $f$ vanishes while $\dot{v}$ remains finite and the inequality holds because when the geodesic enters the horizon, $r$ decreases, so $\dot{r}<0$.  
But the constancy of $E$ then precludes the geodesic from exiting back out of the future horizon, which would require $\dot{r}(\rh)>0$, and therefore $E<0$.  This simple argument implies that case (d) is disallowed, i.e., no geodesic can exit the same future horizon it enters.  

One might try to circumvent the above argument by letting the geodesic come back out through the past horizon instead, as indicated in (e), since there \req{Eonrh} no longer holds as $v\to -\infty$ at the past horizon.  However, this fails for a simpler reason, namely that there would then have to be another turning point for some $r>\rh$, which contradicts our starting assumption that the geodesic reached the horizon all the way from infinity.  Thus case (e) is disallowed as well.

Thus, we have learned that no geodesic with turning point inside the horizon can have both its endpoints anchored to the same boundary.  In other words, {\it no probe geodesic can reach past the horizon}.  Thus from the point of view of asking how deep into the bulk can probe geodesics reach, only Case 2 is relevant.
In particular, we can now WLOG assume that $h(r)>0$ and $f(r)>0$ for all $r$ in the region of interest, i.e.\ in the entire bulk region accessible to probe geodesics.

\paragraph{Case 2 ($\rmin > \rh$):}
The bounce occurs outside the horizon, so it is evident that such a geodesic cannot probe past the horizon.  It is nevertheless interesting to ask how close to the horizon can it approach; in particular, for a given geometry, which types of geodesics have turning point as close as possible to the horizon?

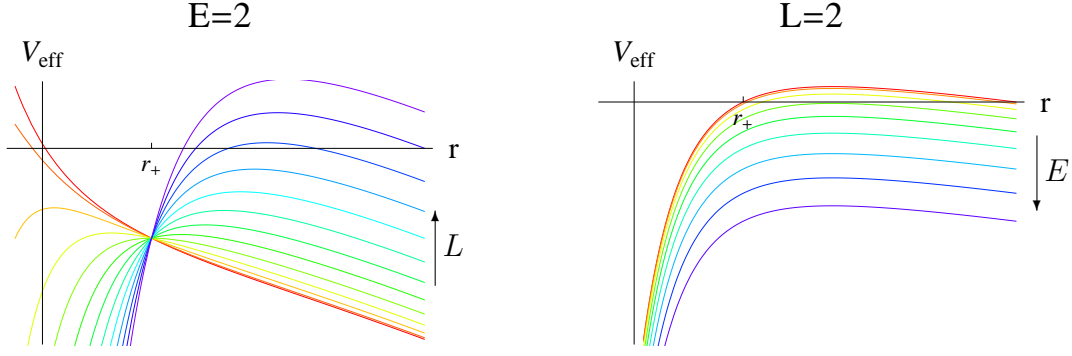
\begin{figure}
\begin{center}
\includegraphics[width=6in]{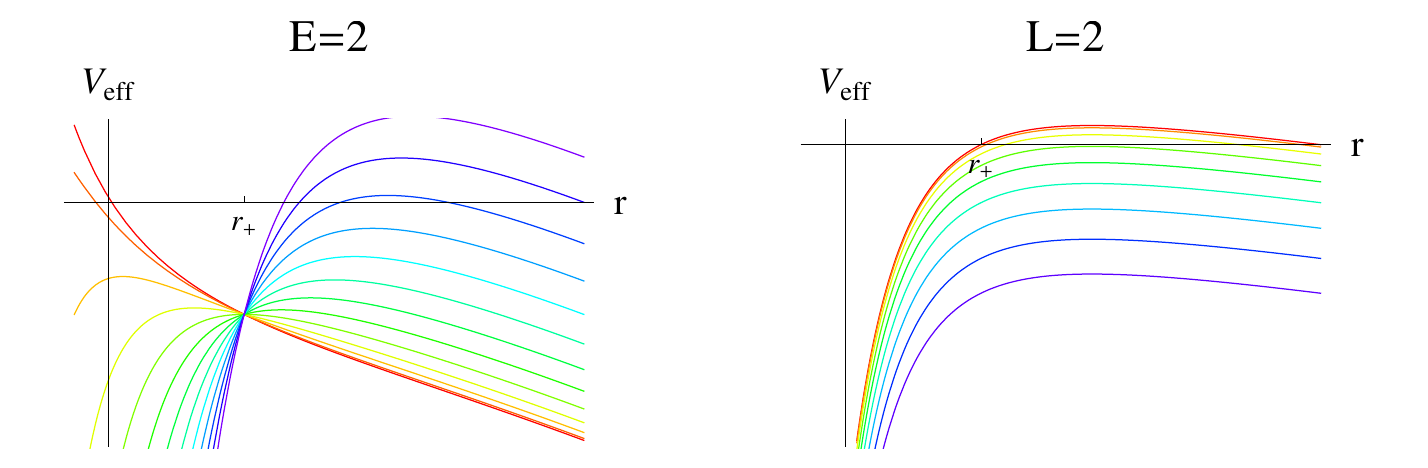}
\begin{picture}(0,0)
\setlength{\unitlength}{1cm}
\put(-9,1){{\vector(0,1){1}}}
\put(-1,3){{\vector(0,-1){1}}}
\put(-8.9, 1.4){$L$}
\put(-0.9, 2.4){$E$}
\end{picture}
\caption{
Effective potentials for global Schwarzschild-AdS$_5$, as $L$ is varied at fixed $E$ (left) and as $E$ is varied at fixed $L$ (right), both for fixed $\rh=1$. 
(left): $L=0,\ldots,3$ in increments of $1/4$ from bottom right (red) to top right (purple) at $E=2$.  (right): $E=0,\ldots,4$ in increments of $1/2$  from top (red) to bottom (purple)  at $L=2$.
}
\label{f:VeffSAdS}
\end{center}
\end{figure}

Though we will briefly comment on null geodesics at the end of this section for completeness, we will primarily focus attention on spacelike geodesics $\kappa = 1$, since outside the horizon this lowers the effective potential at fixed $E,L$ as compared to $\kappa = 0$, and furthermore $E$ and $L$ are more restricted for the null case.  
Let us then consider the two-parameter family of curves $\Veff(r)$ with $\kappa = 1$, each curve parameterized by $E$ and $L$.
From the form of the effective potential, which we illustrate in \fig{f:VeffSAdS} for a prototypical example of Schwarzschild-AdS$_5$, we can see that the curves of $\Veff(r)$ are nested (i.e.\ non-intersecting outside the horizon) if:
\begin{itemize}
\item 
$E$ is fixed and $L$ varies.  In this case $\Veff(r)$ increases as $L^2$ increases for all $r>\rh$, and all curves are pinned to the same value at $r=\rh$.
\item
$L$ is fixed and $E$ varies.  Here $\Veff(r)$ decreases as $E^2$ increases for all $r>0$, and these curves don't intersect even inside the horizon.
\end{itemize}
This suggests that to minimize $\rmin$ along either family of curves, we need to lower $L^2$ or raise $E^2$ respectively until $\Veff$ has local maximum at zero, so that $\Veff(\rmin) = \Veff'(\rmin) = 0$.  This will correspond to unstable circular orbit, and by the above constructive argument exists for a range of energies, each of which determines the requisite angular momentum.  In other words, there will be a one-dimensional curve in $(E,L)$ space along which the corresponding geodesic has a circular orbit.  It then remains to find where along this curve is $\rmin$ minimized.

From \req{Veffrh} and the considerations of the preceding discussion, we see that we can lower $\rmin$ arbitrarily close to $\rh$ by taking $E=0$ (so that $\Veff(\rh)=0$) and adjusting $L$ to the correct value (so that $\Veff'(\rh)=0$).  For general $r$, the latter would be achieved by 
$L^2 = \frac{r^3 \, h'(r)}{2 \, h(r) + r \, h'(r)}$, but at the horizon, this becomes simply
$L^2 =\rh^2$.
Recall that for $E=0$, $\Veff =  \frac{1}{h(r)} \, \left(\frac{L^2}{r^2} - 1\right)$, which has zeros at $r=\rh$ and $r=L$. 
A spacelike geodesic with $E=0$ and $L^2 = \rh^2$ would get trapped in a circular orbit at the horizon, so in order to consider a probe geodesic, we need to take $L^2$ slightly larger.  Nevertheless, this means that we can use probe geodesics to reach arbitrarily close to the horizon.

However, the nearer $\rmin^2 = L^2$ is to $\rh^2$, the longer the geodesic spends in the vicinity of the horizon, so the larger $\Delta \ph $ is.  Moreover, its regularized proper length likewise increases as $L \to \rh$, so such geodesics will be highly subdominant to the one anchored at the same boundary points with $\Delta \ph$ mod $2\pi$.  Restricting $\Delta \ph < 2 \, \pi$ then sets a lower bound on $\rmin$ which lies some {\it finite} distance above the horizon.\footnote{
Note that already for $\Delta \ph >\pi$, there exists a shorter geodesic (on the other side of the black hole) which connects the same boundary points.  This means that the corresponding CFT probe will not have the dominant behaviour determined by this geodesic.
}  

\begin{figure}
\begin{center}
\includegraphics[width=1.9in]{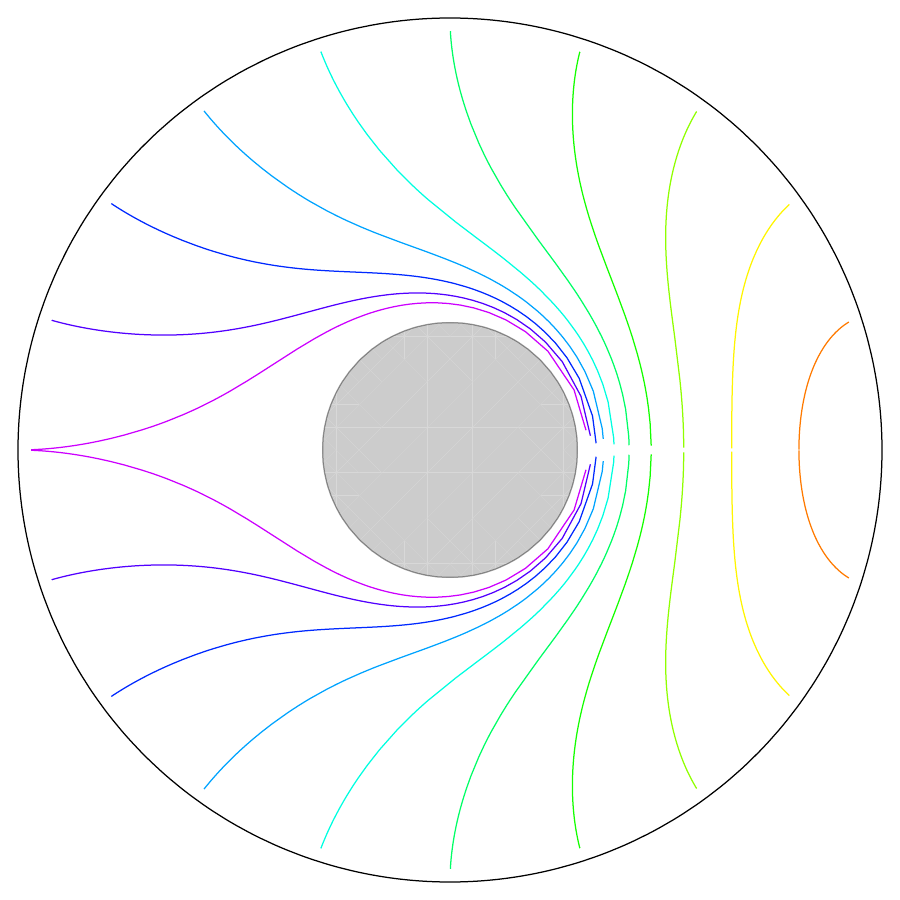}
\hspace{0.2cm}
\includegraphics[width=1.9in]{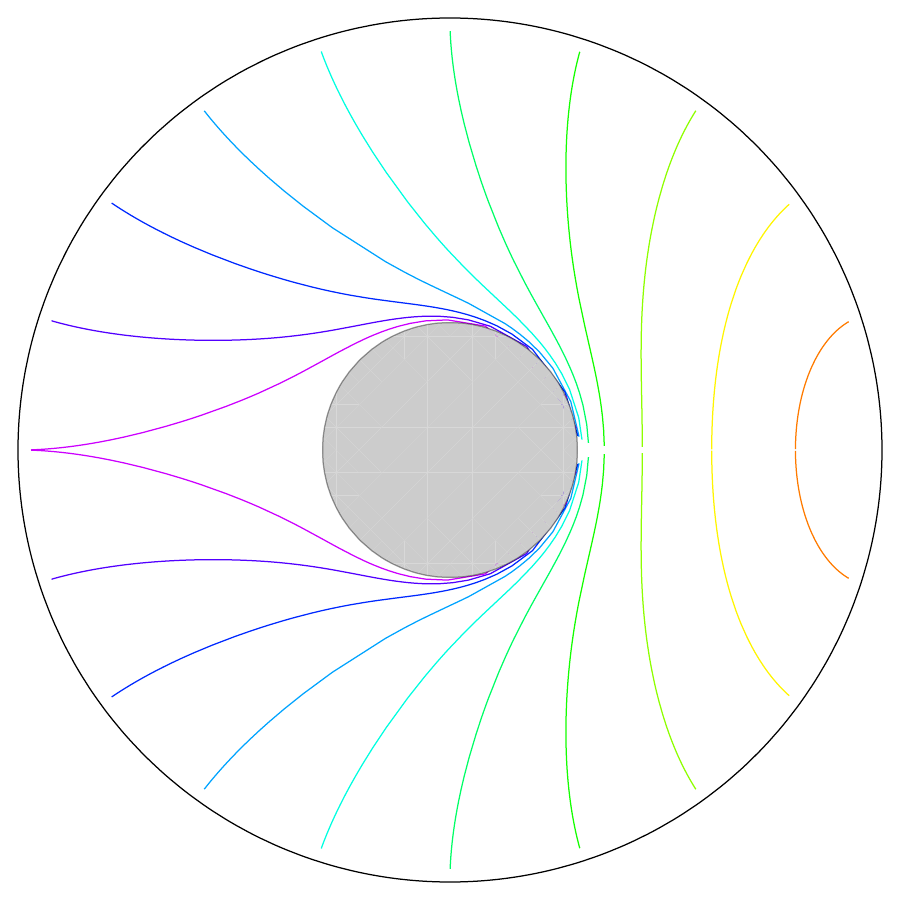}
\hspace{0.2cm}
\includegraphics[width=1.9in]{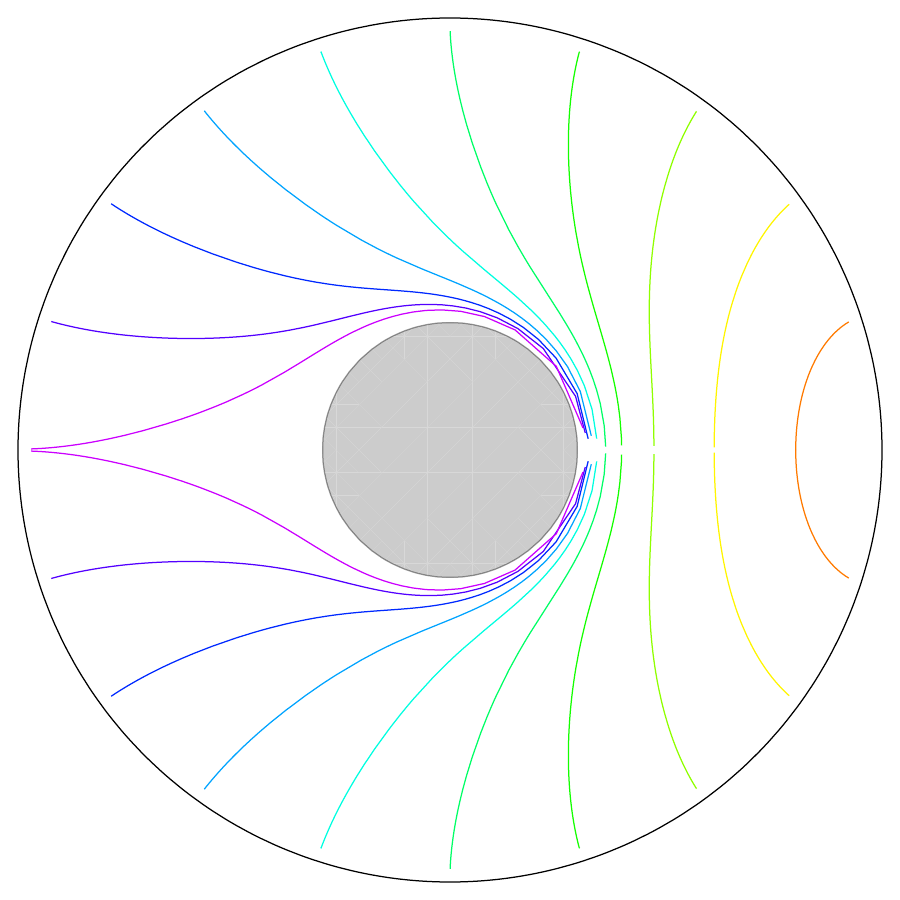}
\caption{
Spacelike $E=0$ geodesics in various backgrounds, for various values of $L$.
We plot the $(r,\ph)$ plane with the radial coordinate given by $\tan^{-1}r$.  The outer circle is the AdS boundary while the inner disk represents a black hole of radius $\rh = 1/2$ in AdS units.  Specific spacetimes used are BTZ (left), Schwarzschild-AdS$_5$ (middle), and extremal Reissner-Nordstrom-AdS$_5$ (right).  The values of angular momenta $L$ are chosen so as to vary $\Delta \ph$ in increments of $\frac{2 \pi}{10}$.  The values of $L$ which gives $\Delta \ph = 2\pi$ (purple curve) are 
$L_{\rm BTZ}= 1.09 \, \rh $, $L_{\rm SAdS} = 1.002  \, \rh$, and $L_{\rm RNAdS} = 1.07 \,  \rh$, respectively.
}
\label{f:BHAdSg}
\end{center}
\end{figure}

To quantify this observation, let us consider several examples of AdS black holes.  In particular, we will use BTZ,  Schwarzschild-AdS$_5$  and extremal Reissner-Nordstrom-AdS$_5$ as prototypical examples.
How close can the probe geodesic with $\Delta \ph \le 2 \pi$ get to the horizon, i.e.\ $\rmin$, of course depends on the specific metric; in particular, from \req{Dtphi} we know that
\begin{equation}
\Delta \ph = 2 \, L \, \int_L^\infty \sqrt{\frac{h(r)}{r^2-L^2}} \, \frac{dr}{r}
\label{}
\end{equation}	
In the case of BTZ where we can obtain the integral in a closed form,
$\Delta \ph = \frac{2}{\rh} \tanh^{-1} \frac{\rh}{L}$, which means that $\Delta \ph = 2\, \pi$ when
\begin{equation}
\frac{\rmin}{\rh} = \frac{L}{\rh} = \frac{1}{\tanh \pi \rh}
\qquad {\rm for \ BTZ} \ .
\label{}
\end{equation}	
From this we see that for very small black holes, the geodesics cannot probe closer to the origin than $\rmin = 1/\pi$ if $\Delta \ph \le 2\, \pi$, whereas for large black holes we can probe exponentially close to the horizon even with this constraint.  
The actual geodesics are plotted in the left panel of \fig{f:BHAdSg} for $\rh =1/2$  and $\Delta \ph$ up to $2\pi$ (which reproduces part of Fig.4 of \cite{Hubeny:2007xt}, where BTZ black hole of various sizes were considered).
For the higher-dimensional AdS black holes, the explicit expression for  $\Delta \ph$ is much more complicated, so we only present the results numerically.  
In the middle and right panels of \fig{f:BHAdSg} we plot geodesics on Schwarzschild-AdS$_5$  and extremal Reissner-Nordstrom-AdS$_5$, respectively. 
Despite the difference in the geometries, the shape of the geodesics varies only mildly.  
We find a qualitatively similar behaviour as in the BTZ case for large black holes, now extending all the way down to the small black hole regime.
However, fixing $\Delta \ph = 2\pi$, the corresponding minimal radii $\rmin = L$ are rather different in the three cases.  In particular,
\begin{equation}
\frac{\rmin - \rh}{\rh} = 
 \left\{ \begin{array}{ccl} 
9.03 \times 10^{-2} &  & \qquad {\rm for \ BTZ} \\
1.85 \times 10^{-3} &  & \qquad {\rm for \ SAdS}_5 \\
6.54 \times 10^{-2} &  & \qquad {\rm for \ RNAdS}_5
\end{array} \right. 
\label{disthor3}
\end{equation}	
The proper lengths also vary in the three cases:  using \req{LpropR}, we find that the regularized proper lengths grow with increasing $\Delta \ph$, reaching the values 3.05, 2.05, and 2.37 when $\Delta \ph = 2\pi$ for BTZ, SAdS, and RNAdS, respectively.

\paragraph{Null geodesics:}
Finally, before ending this section, let us return to our consideration of null geodesics.  
While these cannot reach as far into the bulk as the spacelike geodesics, they may be the most convenient probe to consider from the CFT point of view, thanks to their relation to bulk-cone singularities \cite{Hubeny:2006yu} which are directly accessible in the field theory.

Null geodesics cannot have $E=0$; in fact WLOG we can set $E=1$.  The effective potential can be written as
\begin{equation}
\Veff(r) = \frac{1}{h(r)} \, \left[ \frac{\l^2}{r^2} - \frac{1}{f(r)} \right]
\label{}
\end{equation}	
which  is (strictly) negative on the horizon $\Veff(\rh) <0$.  This means that in order for there to be a bounce, the effective potential has to become positive at some $r>\rh$.  Again, we have a nested family of effective potentials; we can lower $\rmin$ by lowering $\l$ down to a critical value $\l_c$ corresponding to an unstable circular orbit.
The position of this circular orbit is determined by solving $\Veff(r) = \Veff'(r) = 0$, which determines both $\l_c$ and $\rmin$.  Hence, $\rmin$ for null geodesics is typically some finite distance form the horizon and depends only on the geometry -- it cannot be brought closer to the horizon by adjusting the angular momentum.
In the three cases considered above, 
BTZ admits no null circular orbit,
Schwarzschild-AdS$_5$ has null circular orbit at $r_o = \sqrt{2} \, \rh \, \sqrt{1+ \rh^2}$ 
and extremal Reissner-Nordstrom-AdS$_5$ has null circular orbit at $r_o = \sqrt{3} \, \rh \, \sqrt{1+ 2 \, \rh^2}$.   Translating the latter two values into the relative distance from the horizon analogously to that considered in \req{disthor3}, the minimal $\frac{\rmin - \rh}{\rh}$ is $0.58$ and $1.12$ respectively.

Although null geodesics don't have a cost in terms of proper distance, they do have a cost in terms of their endpoints, just as spacelike geodesics do.  Hence bounding $\Delta t$ or $\Delta \ph$ from above gives a greater lower bound on $\rmin$ than that corresponding to the unstable circular orbit.  
As in the spacelike case, though, imposing $\Delta \ph<2\pi$ does not make a substantial difference.

\section{Extremal surface probes}
\label{s:ExtSurf}

Let us now turn our attention from geodesics (which are 1-dimensional extremal `surfaces') to higher dimensional surfaces.  We wish to compare how far into the bulk can these probe. 
For simplicity, we will work with general asymptotically Poincare AdS$_{d+1}$ spacetimes with planar symmetry and time-translation invariance, paying particular heed to black hole spacetimes.  Extremal surfaces in asymptotically global AdS$_{d+1}$ spacetimes with spherical  symmetry and time-translation invariance behave qualitatively similarly, but are algebraically more complicated.  

Before diving in, let us set up the notation and make a few remarks about the choice of coordinates.  In this section we will consider various regions $\CR$ on the boundary.
We will assume that $\CR$  is purely spatial and simply connected.  While the AdS boundary is $d$-dimensional (and therefore has $d-1$ spatial directions) we'll consider regions of varying dimensionality $n=1,2, \ldots, d-1$.  Furthermore, we can consider various geometrical attributes of this region, such as its $n$-dimensional ``area" $A(\CR)$ or its ``extent" $X(\CR)$; we'll define the latter as the maximal geodetic distance between any two points within $\CR$, where all boundary quantities are measured with respect to the boundary Minkowski metric $\eta_{\mu\nu}$.

\begin{figure}
\begin{center}
\includegraphics[width=4in]{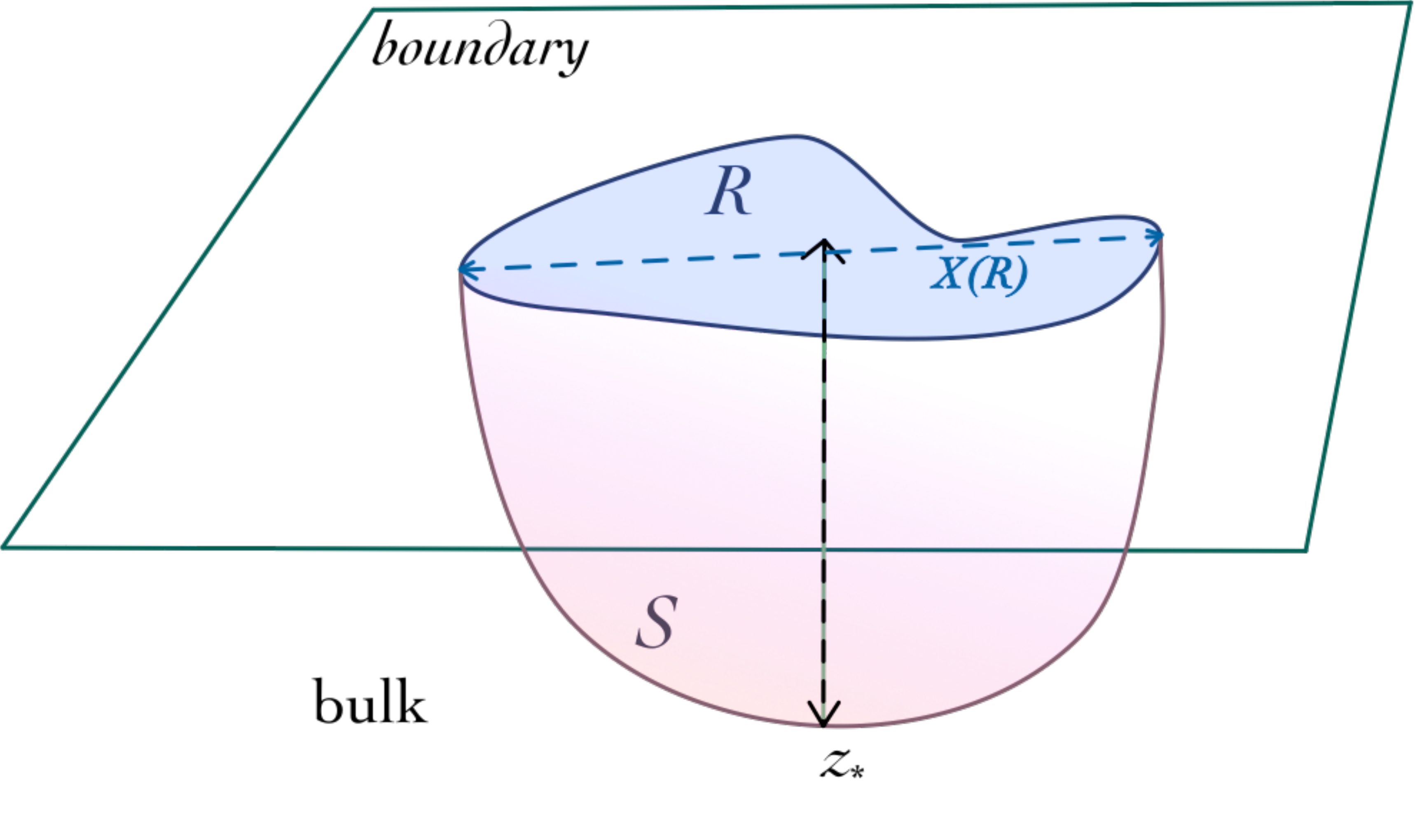}
\caption{
Sketch of the general set-up used in \sec{s:ExtSurf}.  To a given $n$-dimensional region $\CR$ on the boundary we associate an extremal surface $\CS$ in the bulk.  A useful quantity characterizing the surface $\CS$ is its maximal reach $\zmax$ into the bulk, and we will characterize the boundary  regions $\CR$ by their shape, area, or extent $X(\CR)$.
}
\label{f:SurfExpl}
\end{center}
\end{figure}

To a given boundary region $\CR$, we associate a bulk surface $\CS(\CR)$, defined as the extremal area bulk surface anchored on the boundary of $\CR$, i.e.\ $\p {\cal S}= \p \CR$.  
 For orientation, the set-up is sketched in \fig{f:SurfExpl}.
This surface is obtained by extremizing its proper area in the bulk metric.  Although by virtue of extending to the AdS boundary the area of $\CS$ is infinite, the problem of finding the extremal surface is nevertheless well-defined.  In practice when comparing this bulk probe with the corresponding CFT quantity, we regulate to get a finite answer (as described e.g.\ in \cite{Ryu:2006ef}; see also \cite{Myers:2012ed,Liu:2012ee} for more recent discussion).
Since we will restrict our considerations to bulk spacetimes with planar symmetry and time-translation invariance, we can define the bulk radial direction unambiguously (i.e., geometrically), and thereby compare the ``reach" of various surfaces $\CS$ in a natural way.

Having clear definition of the bulk radial direction everywhere in the bulk spacetime (as the unique direction orthogonal to all the other symmetries), all that remains is to find a convenient gauge in which to compare the reach of various surfaces in different spacetimes consistently.  For this task, it is simplest to pick Fefferman-Graham coordinates, in which the bulk metric takes the form
\begin{equation}
ds^2 = \frac{1}{z^2} \, \left[ -g(z) \, dt^2 + k(z) \, dx_i \, dx^i +dz^2 \right]
\label{genPoincmetFG}
\end{equation}	
with $i =1, \ldots, d-1$.
This form is  uniquely defined for any given spacetime by requiring $g_{zz} = 1/z^2$ and $g_{z\mu} = 0$.  Phrased more geometrically, $z$ is related to the affine parameter $\lambda$ (measuring the proper length) along an ingoing radial spacelike geodesic as
$z = e^{\lambda}$.
Note that $z=0$ corresponds to the AdS boundary.
We can now define the ``maximal reach" of $\CS$, $\zmax$ by the largest $z$ value along $\CS$.

In practice, however, \req{genPoincmetFG} is not necessarily the best gauge for writing the equations of motion for the surfaces we will study, because the Lagrangian for $n$-dimensional extremal surface would contain factors of $k(z)^{\frac{n-1}{2}}$.  It will turn out to be simpler for our considerations to use a different radial coordinate, $\znFG$, such that we keep $g_{ii} = 1/z^2$, and instead shift the non-trivial $z$-dependence to $g_{zz}$, as is  more conventional in context of studying black holes:
\begin{equation}
ds^2 = \frac{1}{\znFG^2} \, \left[ -f(\znFG) \, dt^2 +  dx_i \, dx^i 
+ h(\znFG) \, d\znFG^2 \right]
\label{genPoincmet}
\end{equation}	
Note  the use of tilde above $\znFG$ to emphasize the distinction from the Fefferman-Graham coordinates; we'll use this convention throughout the present section.
In both \req{genPoincmetFG} and \req{genPoincmet} the AdS asymptotics requires  the functions $g,k \to 1$ as $z \to 0$ and\footnote{The functions $f$ and $h$ are of course distinct from those used in the global AdS calculations of the previous section.}
$f,h \to 1$ as $\znFG \to 0$.  
For example, the planar Schwarzschild-AdS$_{5}$ spacetime corresponds to
\begin{equation}
f(\znFG) = 1 - \frac{\znFG^4}{\znFGh^4} \ , \quad
h(\znFG) = \frac{1}{1 - \frac{\znFG^4}{\znFGh^4}} \ , \quad
g(z)= \frac{ \left( 1 - \frac{z^4}{\zh^4} \right)^{\! 2}}{1 + \frac{z^4}{\zh^4} } \ ,
\quad {\rm and} \quad
k(z) = 1 + \frac{z^4}{\zh^4}  \ .
\label{planarSAdSfhgk}
\end{equation}	
  In the coordinates of \req{genPoincmet}, we will define the maximal reach of $\CS$ by the largest value of $\znFG$ attained by $\CS$, and denote it by $\znFGmax$.

Whereas it will be easier to find the reach of $\CS$ in terms of $\znFGmax$, we will convert it to $\zmax$ for purposes of comparison between spacetimes.  In particular, to convert between  \req{genPoincmetFG} and \req{genPoincmet}, we use the coordinate transformation (obtained by matching $g_{ii}$)
\begin{equation}
\znFG
	= \frac{z}{\sqrt{k(z)}}
\label{znFGofz}
\end{equation}	
which then allows us to relate the metric functions:
\begin{equation}
h(\znFG) = \frac{1}{k(z)}  \left[ \frac{d}{dz} \left( \frac{z}{\sqrt{k(z)}} \right) \right]^{-2 }
	= \left( 1 - \frac{z}{2} \frac{k'(z)}{k(z)} \right)^{\!  -2}
	\ands
f(\znFG) = \frac{g(z)}{k(z)} \ . 
\label{fkfgRelns}
\end{equation}	
Once we know $k(z)$, we need to invert \req{znFGofz} to find $z(\znFG)$, and hence $\zmax$ from $\znFGmax$.  This  however assumes that $\znFG(z) $ is a strictly monotonically increasing function, which a-priori need not be the case.  Nevertheless, we will now argue that {\it within the entire region of relevance}, i.e.\ where $\CS$ can possibly reach, $\znFG(z) $ does remain monotonic.  The argument relies on a statement whose full proof we leave to \sec{s:ExtSurfHor}, that extremal surfaces can't reach past horizons whilst fully anchored on the boundary.  The fact that outside the horizon, where $\CS$ can  reach, $\znFG(z)$ must be monotonic, can be seen as follows.

Observe that $\znFG(z)$ is monotonic as long as $\frac{d \znFG}{dz} >0 $.  Now suppose this condition is violated at some $z_0$.  In other words, 
$\frac{d}{dz} ( z/\sqrt{k(z)}) \mid_{z_0}= 0$.
Then we see from \req{fkfgRelns} that this means $h(\znFG)$ diverges at the corresponding value of $\znFG(z_0)$. Assuming the full metric is non-singular there, this means that $z=z_0$ describes a null surface, specifically a horizon.  For static spacetimes, this is both a Killing horizon for $\p_t^a$ 
as well as an event horizon for the bulk spacetime.
Our assertion (proved \sec{s:ExtSurfHor}) that extremal surfaces can't reach past  horizons then implies that  our extremal surface $\CS$ cannot reach that far, i.e., $\zmax < z_0$.  Hence within the entire region of relevance for any surface $\CS$, \req{znFGofz} is necessarily monotonic and therefore invertible. 

We now consider specific cases of extremal surfaces $\CS$ anchored on various regions $\CR$ in various bulk geometries, building up from the simplest case in a self-contained manner.
Apart from taking as $\CR$ as an infinite strip or a round ball in pure AdS (which have already been partly considered previously, cf.\ e.g.\ \cite{Ryu:2006ef,Hubeny:2007xt}), we generalize these constructions to arbitrary asymptotically Poincare AdS spacetimes of the form \req{genPoincmet}, giving explicit results for planar Schwarzschild-AdS$_5$ and extremal Reissner-Nordstrom-AdS$_5$.  We end the section by discussing extremal $n$-surfaces in AdS$_{d+1}$ ending on generic regions, to see how different shapes of $\CR$ affect the reach $\zmax$ of $\CS$.  In the next section, we proceed to generalize the set-up even further, and consider arbitrary regions $\CR$ in general asymptotically Poincare AdS spacetimes of the form \req{genPoincmet}, in order to ascertain in full generality that $\CS$ cannot probe past a bulk horizon.

\subsection{Extremal $n$-surface in Poincare AdS$_{d+1}$ ending on a strip}
\label{s:ExtStripPoinc}

\begin{figure}
\begin{center}
\includegraphics[width=3in]{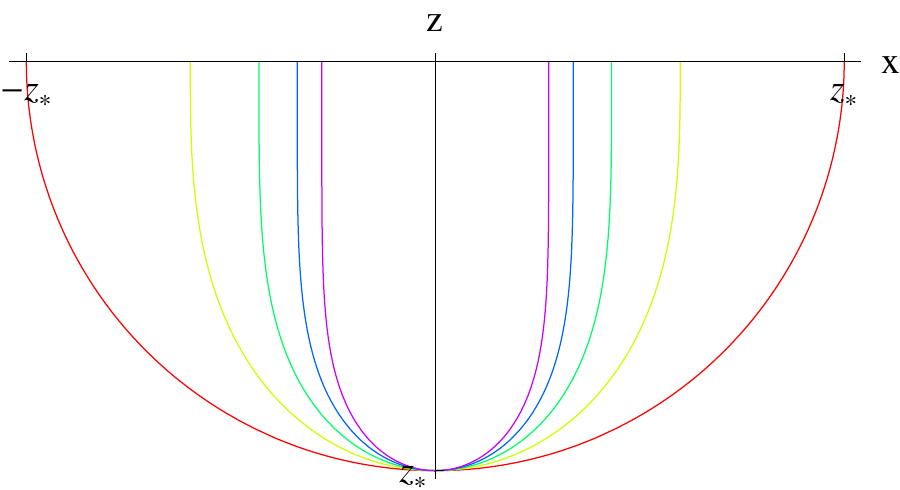}
\hspace{1.5cm}
\includegraphics[width=2.4in]{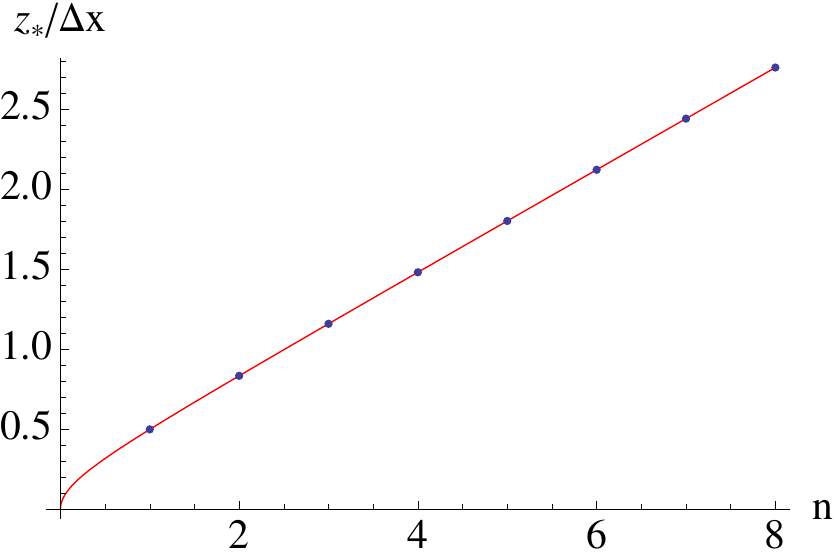}
\caption{
Left: cross-section of $n$-dimensional extremal surfaces in Poincare AdS$_{d+1}$, with varying dimensionality $n=1,2,\ldots,5$: the outermost (red) curve corresponds to $n=1$ while the innermost (purple) curve to $n=5$.  Note that $d$ (as long as it is large enough to accommodate the surface) does not enter.  Right: corresponding ratio of maximal bulk radial extent $\zmax$ to its boundary size $\Delta x$ grows approximately linearly with $n$.
}
\label{f:minsurfAdS}
\end{center}
\end{figure}

Consider Poincare AdS$_{d+1}$, with one direction $x$ singled out, and let us 
define the boundary region $\CR$ to be an $n$-dimensional strip
 specified by 
$x\in [-\frac{\Delta x}{2}, \frac{\Delta x}{2}]$ and $y_i \in (-\infty,\infty)$ for $i=2,\ldots n$.
Note that both the extent and the area of $\CR$ are infinite in this case.
We wish to find the $n$-dimensional extremal surface $\CS$ anchored on $\p \CR$.
The line element \req{genPoincmetFG} with these $n$ directions singled out can be written as
\begin{equation}
ds^2 = \frac{1}{z^2} \, \left[ -dt^2 + dx^2 + \sum_{i=2}^{n} dy_i^2 + \sum_{j=n+1}^{d-1} d\tilde{y}_j^2 + dz^2 \right]
\label{AdSPoincmet} \ .
\end{equation}	
Let us make the gauge choice for the $n$ coordinates $\sigma^a$ on $\CS$ to be simply $\sigma^1 = x$ and $\sigma^i = y^i$ for $i = 2,\ldots n$.  By translational invariance along the $y^i$ directions, $\CS$ will only depend on $x$, so its profile will be specified by the function $z(x)$.  Denoting $\dot{} \equiv \frac{\p}{\p x}$, we calculate the induced metric $G_{ab}$ on $\CS$ in terms of the spacetime metric $g_{\mu\nu}$ and coordinates $X^{\mu}$ in the usual way,
$G_{ab} = g_{\mu\nu} \, \p_a X^{\mu} \, \p_b X^{\nu}$,
finding the determinant 
$G \equiv \det G_{ab} = \frac{1+\dot{z}(x)^2}{z(x)^{2n}}$. 
One can minimize the area of this surface,
\begin{equation}
A = \int \sqrt{G} \, d^{n}\sigma = V^{n-1} \int_{-\frac{\Delta x}{2}}^{\frac{\Delta x}{2}}
\frac{\sqrt{1+\dot{z}(x)^2}}{z(x)^{n}} \, dx \ ,
\label{minsurfA}
\end{equation}	
by solving the corresponding Euler-Lagrange equations
\begin{equation}
z \, \ddot{z} + n \, \dot{z}^2 + n = 0
\label{}
\end{equation}	
whose solution can be written as an inverse Hypergeometric function.
An easier method of obtaining the solution is to note that the Hamiltonian constructed from the Lagrangian
$\CL(z(x),\dot{z}(x)) = \frac{\sqrt{1+ \dot{z}^2}}{z^n}$ is conserved since there is no explicit $x$ dependence, namely
\begin{equation}
\CH = \frac{\p \CL}{\p \dot{z}} \, \dot{z} - \CL = \frac{-1}{z^n \, \sqrt{1+ \dot{z}^2}} 
= \CH(x=0) \equiv \frac{-1}{\zmax^n} \ ,
\label{}
\end{equation}	
which allows us to obtain the explicit expression for $x(z)$ by simply integrating:
\begin{eqnarray}
\pm x(z) &=& \int_{\zmax}^{z} \frac{\tilde{z}^n}{\sqrt{\zmax^{2n}-\tilde{z}^{2n}}} \ d\tilde{z} \cr
&=&  \frac{z^{n+1}}{(n+1) \, \zmax^n} \ {}_2 F_1 
\left[\frac{1}{2}, \frac{n+1}{2n},  \frac{3n+1}{2n}, \frac{z^{2n}}{\zmax^{2n}} \right]
- \zmax \,  \frac{\sqrt{\pi} \, 
\Gamma \left[ \frac{3n+1}{2n} \right]}{(n+1) \,  \Gamma \left[ \frac{2n+1}{2n} \right]} \ .
\label{xofzPoincAdS}
\end{eqnarray}	
Note that this solution automatically incorporates that $x(z=\zmax) = 0$, and using 
$x(z = 0) = \pm \frac{\Delta x}{2}$, we find that the deepest such a surface penetrates, $\zmax$, is related to the spread of the surface, $\Delta x$, by\footnote{
This was already calculated e.g.\ in \cite{Ryu:2006ef}, and subsequently in \cite{Hubeny:2007xt} using slightly different method based on null expansions.
The fact the $\zmax$ grows linearly with $\Delta x$ is guaranteed by conformal invariance.}
\begin{equation}
\zmax = \Delta x \, \frac{n}{\sqrt{\pi}} \, 
\frac{\Gamma \left( \frac{2n+1}{2n} \right) }{\Gamma \left( \frac{n+1}{2n} \right) } \ .
\label{nstripzmax}
\end{equation}	
The set of solutions \req{xofzPoincAdS} is plotted in \fig{f:minsurfAdS}  for various $n$.  It is easy to check that at large $n$, the `distance' $\zmax$ which an $n$-dimensional surface penetrates into the bulk for a fixed strip width $\Delta x$ grows linearly with $n$, 
$\zmax/\Delta x \sim n/\pi$.
The fact that $\zmax/\Delta x$ increases with increasing $n$ can be understood intuitively by noting that the higher the dimensionality, the greater the price to pay for area of the surface near the boundary, so the steeper  the surface in this region becomes, so as to get deeper into the bulk faster.
As we can easily check, at the other limit, $n=1$, we reproduce the well-known result that a spacelike geodesic anchored at $x=\pm \frac{\Delta x}{2}$ (and $t,z,y_i = 0$) is described by the semi-circle
\begin{equation}
x^2 + z^2 = \zmax^2 = \left( \frac{\Delta x}{2}\right)^2 ,
\label{}
\end{equation}	
so the deepest into the bulk that such a geodesic penetrates is given  by $\zmax = \frac{\Delta x}{2}$.

In general, we see that as we increase the dimensionality $n$ of the surface, for a fixed depth $z_{\ast}$ to which such a surface reaches, the strip width $\Delta x$ decreases.  Conversely, keeping $\Delta x$ fixed, higher dimensional surfaces reach deeper into the bulk.  
This observation then seems to suggest the lesson that higher-dimensional surfaces appear to be better probes of the bulk geometry.
As the above explicit results hold for pure Poincare AdS, let us now generalize the geometry slightly, and repeat the calculation to see if the same lesson applies in more general backgrounds.

\subsection{Extremal $n$-surface in asymptotically AdS$_{d+1}$ ending on a strip}
\label{s:ExtStripAsPoinc}

We want to compare the reach $\zmax$ of an extremal surface $\CS$ corresponding to the infinite strip $\CR$ considered in \sec{s:ExtStripPoinc}, in the more general  bulk spacetimes \req{genPoincmetFG}, or more conveniently \req{genPoincmet}. It turns out easier to obtain $\znFGmax$ first and then use  the conversion between the respective bulk radial coordinates given by \req{znFGofz} to obtain $\zmax$.
It will also be convenient to split the $x^i$ directions in the same manner as in \sec{s:ExtStripPoinc} above.  To that end, we write the bulk metric in the form
\begin{equation}
ds^2 = \frac{1}{\znFG^2} \, \left[ -f(\znFG) \, dt^2 + dx^2 + \sum_{i=2}^{n} dy_i^2 
+ \sum_{j=n+1}^{d-1} d\tilde{y}_j^2 + h(\znFG) \, d\znFG^2 \right]
\label{genPoincmetstrip}
\end{equation}	
where we use the tilde on $\znFG$ to emphasize the distinction from the Fefferman-Graham coordinates.

Note that since the spacetime \req{genPoincmetstrip} is static, the extremal surface will lie on constant $t$ spacelike slice of the bulk, so the metric function $f(\znFG)$ will not enter our calculations.
Analogously to \req{minsurfA}, the area of this extremal surface is
\begin{equation}
A = V^{n-1} \int_{-\frac{\Delta x}{2}}^{\frac{\Delta x}{2}}
\frac{\sqrt{1+ h(\znFG) \, \dot{\znFG}(x)^2}}{\znFG(x)^{n}} \, dx
\label{minsurfAf}
\end{equation}	
and we can again use the conservation of Hamiltonian to write the profile $x(\znFG)$ in integral form:
\begin{equation}
\pm x(\znFG) = \int_{\znFGmax}^{\znFG} \frac{\sqrt{h(\bar{z})} \, \bar{z}^n}{\sqrt{\znFGmax^{2n}-\bar{z}^{2n}}} \ d\bar{z} \ .
\label{xofzasPoinc}
\end{equation}	
To proceed further, we have to specify $h(\znFG)$.  However, if the spacetime is smooth and asymptotically AdS, we can Taylor-expand $\sqrt{h(\znFG)}$ around $\znFG=0$ and integrate each term in the expansion.  Defining
\begin{equation}
\sqrt{h(\znFG)} = \sum_{m=0}^{\infty} \sfcf_m \, \znFG^m
\label{sqrtfTaylor}
\end{equation}	
(with $\sfcf_0 = 1$),
we have, analogously to \req{xofzPoincAdS},
\begin{eqnarray}
\pm x(\znFG) &=& \sum_{m=0}^{\infty} \sfcf_m \, \int_{\znFGmax}^{\znFG} \frac{\bar{z}^{n+m}}{\sqrt{\znFGmax^{2n}-\bar{z}^{2n}}} \ d\bar{z} \cr
&=&  \frac{\znFG^{n+1}}{\znFGmax^n}  \sum_{m=0}^{\infty} \frac{\sfcf_m \, \znFG^m}{n+m+1} 
	\ {}_2 F_1 \left[\frac{1}{2}, \frac{n+m+1}{2n},  \frac{3n+m+1}{2n}, 
	\frac{\znFG^{2n}}{\znFGmax^{2n}} \right] \cr
&&\qquad
- \sqrt{\pi} \, \znFGmax \,  \sum_{m=0}^{\infty}\sfcf_m \, \znFG^m \,\frac{1}{m+1} 
\frac{\Gamma \left[ \frac{n+m+1}{2n} \right]}{\Gamma \left[ \frac{m+1}{2n} \right]} \ .
\label{xofzasPoincAdS}
\end{eqnarray}	

We can now read off the width of the strip $\Delta x_n$ for the $n$-dimensional extremal surface in terms of $\znFGmax$ and $\sfcf_m$:
\begin{equation}
\Delta x_n = \frac{\sqrt{\pi}}{n}   \sum_{m=1}^{\infty}\sfcf_{m-1} \, \znFGmax^m \,
\frac{\Gamma \left[ \frac{n+m}{2n} \right]}{\Gamma \left[ \frac{2n+m}{2n} \right]} \ .
\label{DxasPoinc}
\end{equation}	
It is easy to see that in the large-$n$ limit, 
\begin{equation}
\Delta x_n \sim  \znFGmax \, \sqrt{h(\znFGmax)} \ \frac{\pi}{n} \qquad {\rm as } \ n \to \infty \ ,
\label{DxasPoincLargen}
\end{equation}	
which indeed demonstrates monotonicity with $n$ in the large $n$ regime for any spacetime of the form \req{genPoincmetstrip}.  However, in our set-up $n\le d-1$, so it's more convenient to examine \req{DxasPoinc} for small $n$ explicitly by evaluating the Gamma functions.
The ratio of the Gamma functions in \req{DxasPoinc} scales as $\sim \sqrt{1/m}$ at large $m$, however these coefficients are slightly larger than 1 for small $m$, so we cannot straightforwardly use \req{DxasPoincLargen} as an upper bound for any $n$.  To build intuition, let us then simplify the problem further.

To this end, let us consider just the first subleading term in the expansion \req{sqrtfTaylor}.   In particular, let $\sqrt{h(\znFG)} =1+ \sfcf_d \, \znFG^d$.
Then $\Delta x_n$ in \req{DxasPoinc} consists of only two terms, and we can check at which value of $\sfcf_d \, \znFGmax^d$ does $\Delta x_n$ become smaller than $\Delta x_m$ for some $m<n$.  It turns out that this value is always negative, and moreover it also necessarily renders $\Delta x_n$ negative.\footnote{
The former is easy to see; the latter, although somewhat tedious, can likewise be verified algebraically using monotonicity properties of the Gamma functions involved. 
}  This suggests that in the physical regime, $\Delta x_n$ is indeed monotonically decreasing with $n$ over the full range of $n$, and of course increasing with $\znFGmax$; so conversely, if we fix $\Delta x$, then $\znFGmax$ increases with increasing $n$ in this larger class of bulk geometries, just as it did for pure AdS.

Having motivated our expectations, we now argue this in far greater generality by considering the integral form of $\Delta x$ given by \req{xofzasPoinc}, which we can rewrite more suggestively as
\begin{equation}
\Delta x = 2\, \int_{0}^{\znFGmax} \frac{\sqrt{h(\znFG)}}{\sqrt{(\znFGmax/\znFG)^{2n}-1}} \ d\znFG \ .
\label{DxintasPoinc}
\end{equation}	
For fixed $\znFGmax$ (and fixed $h(\znFG)>0$), the denominator in the integrand is larger for larger $n$ when $\znFG\in (0,\znFGmax)$, which implies that the full integral is smaller for larger $n$.  This proves that $\Delta x/\znFGmax$ decreases with $n$, so that at fixed $\Delta x$, $\znFGmax$ increases with $n$ -- i.e., higher-dimensional surfaces probe deeper.

\paragraph{Pure AdS yields largest $\zmax/\Delta x_n$:}
Now that we have analyzed the effect of varying $n$ in a fixed spacetime, let us fix $n$ and instead consider the effect of varying the spacetime, i.e., explore the effect that the deformation $h(\znFG)$ in \req{genPoincmetstrip} has on the extremal surfaces.  
From \req{DxintasPoinc}, we immediately see that if $h(\znFG)\ge1$ $\forall \  \znFG\in (0,\znFGmax)$, then the corresponding $\Delta x$ is greater than the pure AdS ($h(\znFG)=1$) value, for any $n$.  As we see from \req{planarSAdSfhgk}, the planar Schwarzschild-AdS black hole certainly satisfies $h(\znFG)\ge1$, and we expect more generally that this condition would hold for physically sensible situations -- in other words, we expect that any physically admissible matter would create a gravitational potential well.  This is easy to see asymptotically.
Let us again start by considering just the leading order correction to pure AdS, $\sqrt{h(\znFG)} =1+ \sfcf_d \, \znFG^d$.  (Note that from \req{DxasPoinc} we immediately recover that  for any $n$, if $\sfcf_d>0$, $\Delta x_n$ at fixed $\znFGmax$ is greater than the corresponding value in pure AdS.)  So all that remains is to establish the physically relevant sign of $\sfcf_d$.

The physically relevant sign of  $\sfcf_d$ is the one which yields a physically sensible stress tensor on the boundary, namely one with non-negative energy density, pressure, etc..  So in order to determine which class of bulk deformations is physically sensible, we will first extract the boundary stress tensor.  One can do this by several methods; e.g. we can use the Balasubramanian-Kraus construction \cite{Balasubramanian:1999re} in the present coordinates, or we can use the de Haro et.al.'s prescription \cite{deHaro:2000xn} to read off the stress tensor from the metric as expressed in Fefferman-Graham coordinates.

Let us illustrate this in 4+1 bulk dimensions.  
Writing the metric in Fefferman-Graham coordinates as
\begin{equation}
ds^2 = \frac{1}{z^2} \left[ dz^2 +
\left( g^{(0)}_{\mu\nu} + z^2 \, g^{(2)}_{\mu\nu} + z^4 \, g^{(4)}_{\mu\nu} + \ldots \right)
dx^{\mu} \, dx^{\nu} \right]
\label{}
\end{equation}	
with $g^{(0)}_{\mu\nu} = \eta_{\mu\nu}$ implies that $g^{(2)}_{\mu\nu} = 0$, and
$g^{(4)}_{\mu\nu} \propto \langle T_{\mu\nu} \rangle$, where $  \langle T_{\mu\nu} \rangle$ is the expectation value of the boundary stress tensor.
Letting $T_{\mu\nu}$ take the perfect fluid form,
\begin{equation}
T_{\mu\nu}= \rho \, u_\mu \, u_\nu + P \, \left(  \eta_{\mu\nu} + u_\mu \, u_\nu  \right)
\label{}
\end{equation}	
gives
\begin{equation}
ds^2 = \frac{1}{z^2} \left[
 - \left( 1 - \rho \, z^4 \right) \, dt^2 +  \left( 1 +P \, z^4 \right) \, dx_i \,  dx^i +  dz^2 
 \right]  \ .
\label{genFGmetrhoP}
\end{equation}	
For example, for  the planar Schwarzschild-AdS black hole,  expanding the metric \req{planarSAdSfhgk} to $\CO(z^4)$ gives the known values
\begin{equation}
\rho = \frac{3}{\zh^4} = 3 \, \pi^4 \, T^4 \ , \qquad
P = \frac{1}{\zh^4} = \pi^4 \, T^4  = \frac{\rho}{3} \ .
\label{}
\end{equation}	
We invert \req{znFGofz} to relate the Fefferman-Graham  form \req{genFGmetrhoP} to the form \req{genPoincmet} used above, specifically
\begin{equation}
\znFG^2 =  \frac{z^2}{1+ P \, z^4}
\iffs
z^2 =  \frac{1 - \sqrt{1-4 \, P \, \znFG^4}}{2 \, P \, \znFG^2} \ .
\label{ztoznFGcx}
\end{equation}	
Keeping only terms up to $\CO(z^4)$, we can expand \req{genPoincmet} as
\begin{equation}
ds^2 = \frac{1}{\znFG^2} \, \left[
 - (1 - [\rho+P] \, \znFG^4) \, dt^2 + dx_i \, dx^i
 +  (1 + 4 \, P \, \znFG^4)  \, d\znFG^2 \right] 
\label{}
\end{equation}	
which allows us to identify 
\begin{equation}
\sqrt{f(\znFG)} =1+ \sfcf_4 \, \znFG^4 = 1 + 2 \, P \, \znFG^4  \ .
\label{}
\end{equation}	
In particular, this means that the physically sensible sign of $\sfcf_4 =2 \, P$ is indeed positive.

We have now established that asymptotically, a physically sensible deformation of AdS requires the leading correction to $h(\znFG)$ to be positive.  Moreover, as exemplified by the black hole metric \req{planarSAdSfhgk} and argued more generally in \sec{s:ExtSurfHor}, $h(\znFG)$ has to get large near the horizon of any black hole.  Hence what remains unknown is the region between the horizon and the asymptopia.  

We  conjecture that any non-pathological bulk spacetime must satisfy the property 
$h(\znFG)\ge1$ for all $z$ between the horizon and the boundary.
The diagnostic of the pathology involved in this condition being violated would presumably involve unphysical behaviour of higher point functions of the boundary stress tensor, and would be an interesting direction to explore further.
If valid, this statement is sufficient (though not necessary) to prove that in any physically sensible bulk spacetime which is distinct form AdS, an extremal surface anchored on a strip cannot reach as far in the bulk as it could in pure AdS. 

\paragraph{Extremal surfaces can't reach past horizons:}
Given the conclusion of the preceding paragraph, one can ask what happens when the deformation $h(\znFG)$ to AdS takes an extreme form, i.e.\ when the bulk spacetime has a black hole.  Although as already advertised, in \sec{s:ExtSurfHor} we will prove that  extremal surfaces cannot extend past the horizon, it is instructive in this simple set-up to reach the same conclusion using a different argument.
 
Making use of the symmetries of the present set-up, we can rewrite \req{xofzasPoinc} in the form of an effective potential for the surface $\znFG(x)$:
\begin{equation}
\dot{\znFG}^2 + \Veff(\znFG) = 0 \ , \ \qquad {\rm where } \quad
\Veff(\znFG) = \frac{1}{h(\znFG)} \, \frac{\znFG^{2n}- \znFGmax^{2n}}{\znFG^{2n}} \ .
\label{}
\end{equation}	
The effective potential has two zeros  (which may or may not be distinct): one at $\znFG = \znFGmax$, and one at $\znFG = \znFGh$, since  $h$ diverges at the horizon.  Moreover, assuming the horizon is non-degenerate,  $h(\znFG)$ is positive outside the horizon ($\znFG<\znFGh$) but negative inside ($\znFG>\znFGh$), which renders $ \Veff(\znFG)$ positive for all $\znFG$ between $\znFGh$ and $\znFGmax$.  This region is therefore inaccessible to our extremal surface.  This means that we must have $\znFGmax < \znFGh$, showing that the extremal surface cannot penetrate past  the horizon.
(Note that if $\znFGmax = \znFGh$, then $\dot{\znFG}=0$, so the surface hugs the horizon and hence cannot be anchored at the boundary.)

\paragraph{Example for planar Schwarzschild-AdS:}
Let us see the arguments of the preceding paragraph realized explicitly in the case of particular interest, namely the planar Schwarzschild-AdS$_{d+1}$ black hole geometry.  We will consider the non-trivial but algebraically simplest case of $d=4$.  The metric was given in \req{planarSAdSfhgk}, namely:
\begin{eqnarray}
ds^2 
&=& \frac{1}{\znFG^2} \, \left[ -\left( 1 - \frac{\znFG^4}{\znFGh^4}\right) \, dt^2 +  dx_i \, dx^i 
+ \frac{1}{1 - \frac{\znFG^4}{\znFGh^4}} \, d\znFG^2 \right] \cr
&=& \frac{1}{z^2} \left[
 - \frac{ \left( 1 - \frac{z^4}{\zh^4} \right)^2}{ \left( 1 + \frac{z^4}{\zh^4} \right)} \, dt^2 +   \left( 1 + \frac{z^4}{\zh^4} \right)\, dx_i \,  dx^i +  dz^2 
 \right]
\label{planarSAdSle}
\end{eqnarray}	
where $\znFGh$ and $\zh$ correspond to the horizon radius, and are related by $\zh^2 = 2 \, \znFGh^2$.

We can find $\znFGmax$ for an $n$-dimensional surface anchored on a strip of width $\Delta x$ by solving 
\begin{equation}
\Delta x = 2 \, \int^{\znFGmax}_{0} \frac{\sqrt{h(\bar{z})} \, \bar{z}^n}{\sqrt{\znFGmax^{2n}-\bar{z}^{2n}}} \ d\bar{z} 
\qquad {\rm with} \qquad
h(\bar{z}) = \frac{\bar{z}_+^4}{\bar{z}_+^4 - \bar{z}^4}
\label{DxofzasPoinc}
\end{equation}	
for $\znFGmax$.  Although a closed-form expression can be readily obtained only for $n=1$, we can easily compute the behaviour numerically, and confirm that for any $n$, $\Delta x$ diverges as $\znFGmax \to \znFGh$.
To plot the full extremal surfaces, we solve \req{xofzasPoinc} to get $x(\znFG)$, and then invert  \req{znFGofz},
\begin{equation}
\frac{z}{\zh} 
	= \frac{\znFGh}{\znFG} \, 
	\sqrt{1 - \sqrt{ 1 - \left( \frac{\znFG}{\znFGh} \right)^4 }} \ ,
\label{zofznFGSAdS}
\end{equation}	
to find $z(x)$ in Fefferman-Graham coordinates.  
\begin{figure}
\begin{center}
\includegraphics[width=5in]{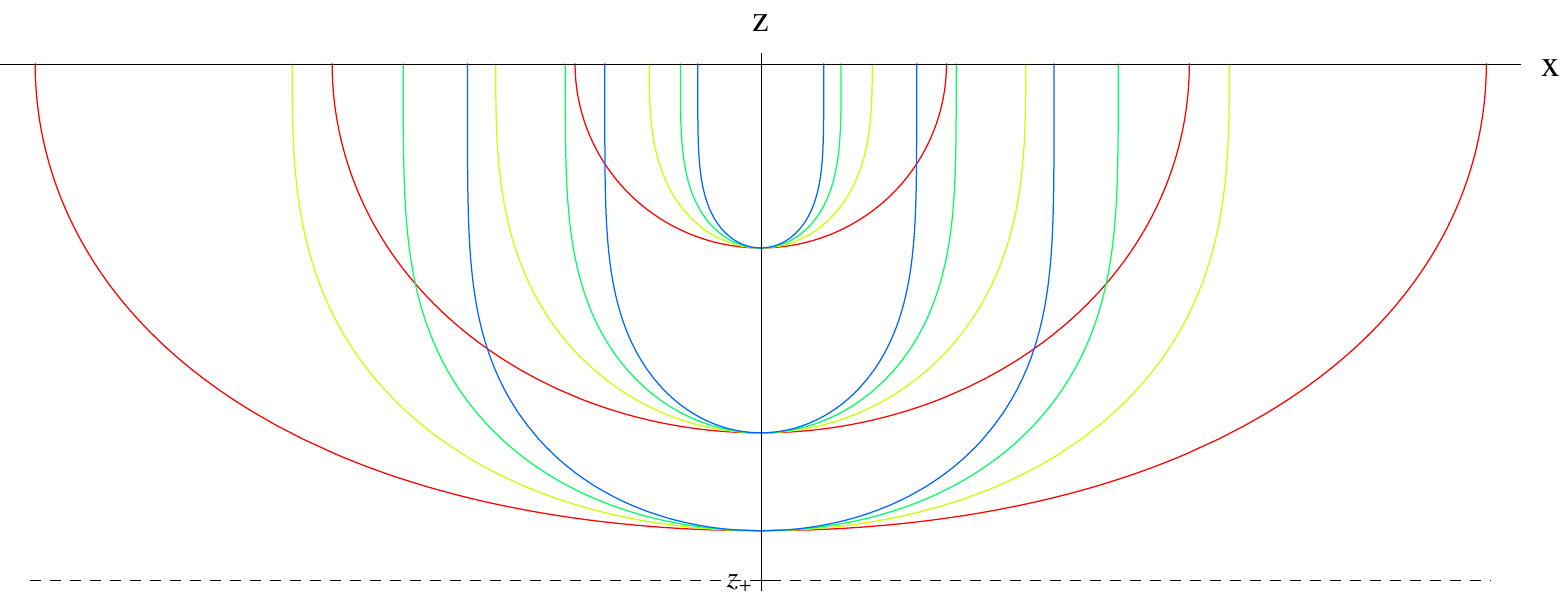}
\caption{
Cross-sections of $n$-dimensional extremal surfaces anchored on a strip in a planar Schwarzschild-AdS$_5$ background \req{planarSAdSle}, for various $n$ and various $\zmax$.  The black dashed line at the bottom corresponds to the horizon. 
The three sets of surfaces, from top to bottom, have $\frac{\znFGmax}{\znFGh} = 0.5,0.9$, and $0.99$, respectively.
 As in \fig{f:minsurfAdS}, for each $\zmax$, the outermost (red) curve corresponds to $n=1$ while the innermost (blue) curve to $n=4$.
}
\label{f:minsurfSAdS}
\end{center}
\end{figure}
The resulting surfaces are plotted in \fig{f:minsurfSAdS}.  The figure shows three sets of $n$ dimensional extremal surfaces, each with\footnote{
Although for $d=4$, only $n < 4$ surfaces are relevant, we include the $n=4$ case to illustrate the pattern more clearly, and as an indicator of absolute upper bound on $\zmax/\Delta x$.
} $n=1,2,3,4$ (color-coded by $n$, with red corresponding to $n=1$), pinned to a given value of $\zmax$; from top set to bottom set, $\znFGmax/\znFGh = 0.5, 0.9$ and $0.99$ (which, using \req{zofznFGSAdS}, translates to $\zmax/\zh \approx 0.36, 0.71$, and $0.90$).
We see that as we change $n$, we encounter similar behaviour as seen for pure AdS in \fig{f:minsurfAdS}:  the higher-dimensional surfaces are `steeper' and therefore have greater $\zmax/\Delta x$.  The event horizon is indicated by the dashed black line at the bottom.  We can clearly see that as $\zmax$ approaches the horizon, the extremal surface gets deformed so as to remain above the horizon.  While the top set of curves ($\znFGmax = \znFGh/2$) are still largely unaffected by the presence of the black hole, the bottom set is noticeably widened, as we predicted.  One can also repeat the analysis for higher-dimensional black holes, which yields qualitatively similar results.

\paragraph{Example for extremal planar Reisser-Nordstrom-AdS:}
The above example illustrated the properties for non-extremal black holes; but before leaving this section, it is interesting to examine what happens for geometries with extremal (degenerate) horizons, since these are qualitatively different.  Most importantly, in such cases there is an infinite proper distance to the horizon along spacelike geodesics, which means that the Fefferman-Graham coordinate $z$ diverges at the horizon.  For this reason, we will examine the extremal surfaces in both sets of coordinates.

\begin{figure}
\begin{center}
\includegraphics[width=5in]{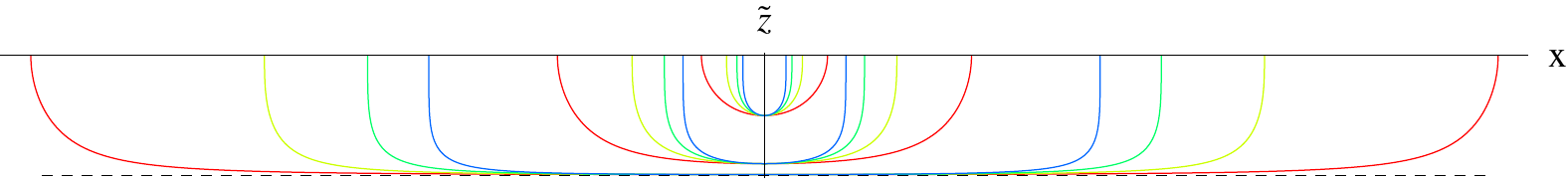} \\
\vspace{1cm}
\includegraphics[width=5in]{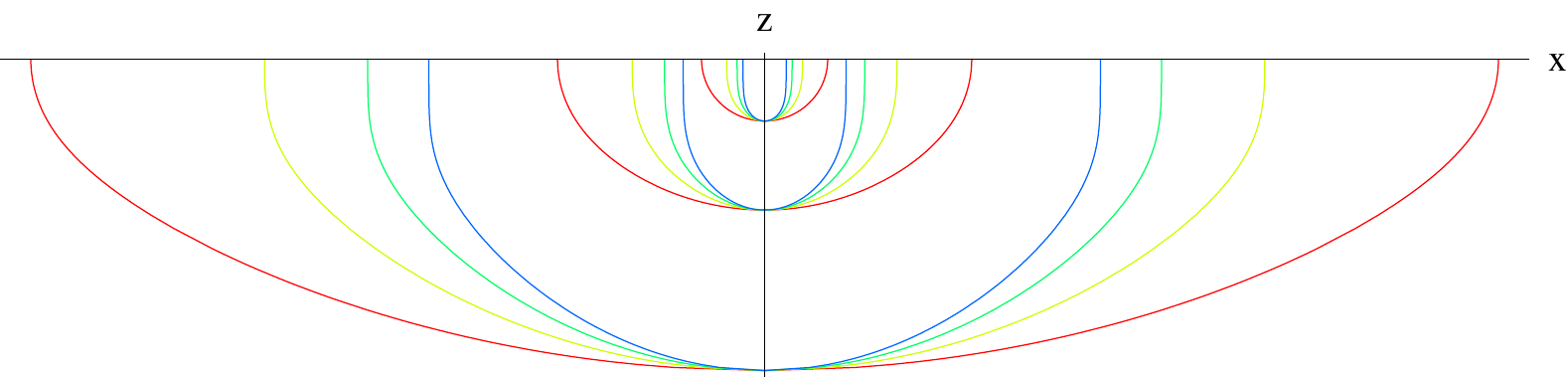}
\caption{
Cross-sections of $n$-dimensional extremal surfaces anchored on a strip in a planar extremal Reisser-Nordstrom-AdS$_5$ background \req{planarRNAdS}, for various $n$ and various $\zmax$.  The black dashed line at the bottom corresponds to the horizon.  As in \fig{f:minsurfAdS}, for each $\zmax$, the outermost (red) curve corresponds to $n=1$ while the innermost (blue) curve to $n=4$.
}
\label{f:minsurfRNAdS}
\end{center}
\end{figure}

As a convenient example, let us take the metric of a 5-dimensional Reissner-Nordstrom-AdS black hole, and use extremality to express it in terms of the horizon radius;
in the form of
\req{genPoincmet} we have
\begin{equation}
f(\znFG) = \frac{1}{h(\znFG)} = 1-m \, \znFG^4 + q \, \znFG^6
	\to \left( 1 - \frac{\znFG^2}{\znFGh^2} \right)^{\! \!  2} \, 
	\left( 1 +2  \frac{\znFG^2}{\znFGh^2} \right) \ .
\label{planarRNAdS}
\end{equation}	
Setting $\zeta \equiv \frac{\znFG}{\znFGh}$ to simplify notation, the conversion to the Fefferman-Graham form  is
\begin{equation}
z = c \, \frac{\zeta}{1+\sqrt{1+2\zeta^2}} \, 
	\left( \frac{2+\zeta^2 + \sqrt{ 3 + 6 \, \zeta^2}}
		{1-\zeta^2} \right)^{\! \! \frac{\sqrt{3}}{6}}
\label{zofznFGRN}
\end{equation}	
with $c=2 \, (2+\sqrt{3})^{\sqrt{3}/6}$.  
This is linear as $\zeta\to 0$, but diverges as $\zeta \to 1$.
As before, we solve \req{xofzasPoinc} with $h$ given by \req{planarRNAdS} to find the extremal surface as $x(\znFG)$, which is shown in the top panel of \fig{f:minsurfRNAdS} for the same parameters as used in \fig{f:minsurfSAdS}.
In these coordinates the surfaces hug the horizon even more sharply than the corresponding surfaces in the analogous plot (not shown) for Schwarzschild-AdS black hole.  Using the conversion \req{zofznFGRN}, in the bottom panel of \fig{f:minsurfRNAdS} we plot the corresponding surfaces in the Fefferman-Graham coordinates.
As advertised, the horizon has now receded infinitely far down, so the surfaces don't cluster; in fact, very close to the horizon they would behave in a self-similar fashion, at fixed $n$ being just rescaled versions of each other.  Nevertheless, these surfaces still get elongated along the $x$ direction due to the presence of the horizon.

\subsection{Extremal $n$-surface in Poincare AdS$_{d+1}$ enclosing a ball}
\label{s:ExtBallPoinc}

Above we have argued that for extremal surfaces anchored on `strips',  higher-dimensional surfaces are better probes of the bulk geometry, in the sense that for a fixed strip width, an $n$-dimensional extremal surface reaches deeper into the bulk for higher $n$.  However, this argument is a bit too glib.  In particular, we have only considered the extent in the $x$ direction and ignored the fact that the higher dimensional surfaces utilize extra directions of infinite extent.  A slightly better comparison would  therefore involve a boundary region of finite extent in all directions.  The most natural such region is a $(d-1)$-ball (with some specified radius $R_0$) in $\RR^{d-1}$.  Let us therefore compare extremal $n$-surfaces anchored on $S^{n-1}$ of radius $R_0$ (i.e.\ with extent $X(\CR)=2R_0$), for different values of $n \le d-1$.  

Let us again start with pure Poincare AdS$_{d+1}$ as a warm-up.
It is convenient to use coordinates adapted to the spherical symmetry of the bounding region, so we re-write \req{AdSPoincmet} as
\begin{equation}
ds^2 = \frac{1}{z^2} \, \left[ -dt^2 + dr^2 + r^2 \, d\Omega_{n-1}^2+ \sum_{j=n+1}^{d-1} d\tilde{y}_j^2 + dz^2 \right]
\label{AdSPoincmetSpher}
\end{equation}	
and choose $\sigma^1 = r$ and $\sigma^i$ with $i = 2,\ldots,n$ to be given by the angles of the $(n-1)$-sphere.  Analogously to \req{minsurfA},
the area of the bulk surface is  then
\begin{equation}
A = \Omega_{n-1}  \, \int_0^{R_0}
	\frac{\sqrt{1+\dot{z}(r)^2}}{z(r)^{n}} \, r^{n-1} \, dr
\label{AzrPoinc}
\end{equation}	
where $\Omega_{n-1} = \frac{2 \pi^{n/2}}{\Gamma(n/2)}$ is the volume of the unit $S^{n-1}$ and $\dot{z} \equiv \frac{d z}{dr}$.  
Unlike in the case of the strip \req{minsurfA}, the Lagrangian now depends explicitly on $r$, so we can't use conservation of Hamiltonian to write the solution as an integral.  Instead, we obtain the equation of motion from the Euler-Lagrange equations,
\begin{equation}
\ddot{z} + (1 + \dot{z}^2 ) \, \left[\frac{n}{z} + \frac{(n-1)}{r} \, \dot{z} \right] =  0 \ .
\label{ELzrPoinc}
\end{equation}	
Although the equation of motion for such a surface depends on $n$, it can be easily verified that this equation is in fact satisfied by a very simple solution
\begin{equation}
z(r) = \sqrt{R_0^2 - r^2}
\label{}
\end{equation}	
for all $n$ -- that is, the minimal surface is simply an $n$-hemisphere, whose bulk extent 
$\zmax = R_0$ is independent of $n$.  

Hence we learn that for round regions $\CR$, it is no longer the case that higher-dimensional surfaces would necessarily probe deeper, which advocates caution in interpreting the lesson of the strip and warrants further exploration.

\subsection{Extremal $n$-surface in asymptotically AdS$_{d+1}$ enclosing a ball}
\label{s:ExtBallAsPoinc}

In the previous subsection, we have seen that all $n$-dimensional extremal surfaces anchored on a $n$-ball of radius $R_0$ on the boundary reach the same distance $\zmax = R_0$ into the bulk.  However, it is clear that this can happen only due to a special cancellation in the bulk geometry:  both the geometry and the boundary region of interest had high degree of symmetry.

We leave the examination of what happens under deformation of the shape of $\CR$ to \sec{s:ExtGenPoinc}; here we instead take a brief detour into addressing the effect of deforming the spacetime while keeping $\CR$ fixed, analogously to the considerations of \sec{s:ExtStripAsPoinc}.  In particular, we will consider arbitrary  spacetimes of the form \req{genPoincmet}, characterized by two arbitrary functions $f(\znFG)$ and $h(\znFG)$.

The bulk metric adapted to the region of interest $\CR$ is
\begin{equation}
ds^2 = \frac{1}{\znFG^2} \, \left[ - f(\znFG) \, dt^2 + dr^2 + r^2 \, d\Omega_{n-1}^2+ \sum_{j=n+1}^{d-1} d\tilde{y}_j^2 + h(\znFG) \, d\znFG^2 \right]
\label{AdSPoincmetSpher}
\end{equation}	
which gives the area of the corresponding bulk extremal surface $\CS$,
\begin{equation}
A = \Omega_{n-1}  \, \int_0^{R_0}
	\frac{\sqrt{1+ h(\znFG) \, \dot{\znFG}(r)^2}}{\znFG(r)^{n}} \, r^{n-1} \, dr \ .
\label{}
\end{equation}	
The corresponding Euler-Lagrange equations in this case are more complicated,
\begin{equation}
h(\znFG) \, \ddot{\znFG} + (1 + h(\znFG) \, \dot{\znFG}^2 ) \, \left[\frac{n}{\znFG} 
+ \frac{(n-1)}{r} \, h(\znFG) \, \dot{\znFG} \right] + \frac{h'(\znFG)}{2} \, \dot{\znFG}^2 
=  0
\label{}
\end{equation}	
but can be solved numerically.  

For illustrative purposes, we again consider the planar Schwarzschild-AdS$_5$ black hole metric \req{planarSAdSle}, and solve for the radial profile of $\CS$, using \req{zofznFGSAdS} to convert to Fefferman-Graham coordinates.
\begin{figure}
\begin{center}
\includegraphics[width=5in]{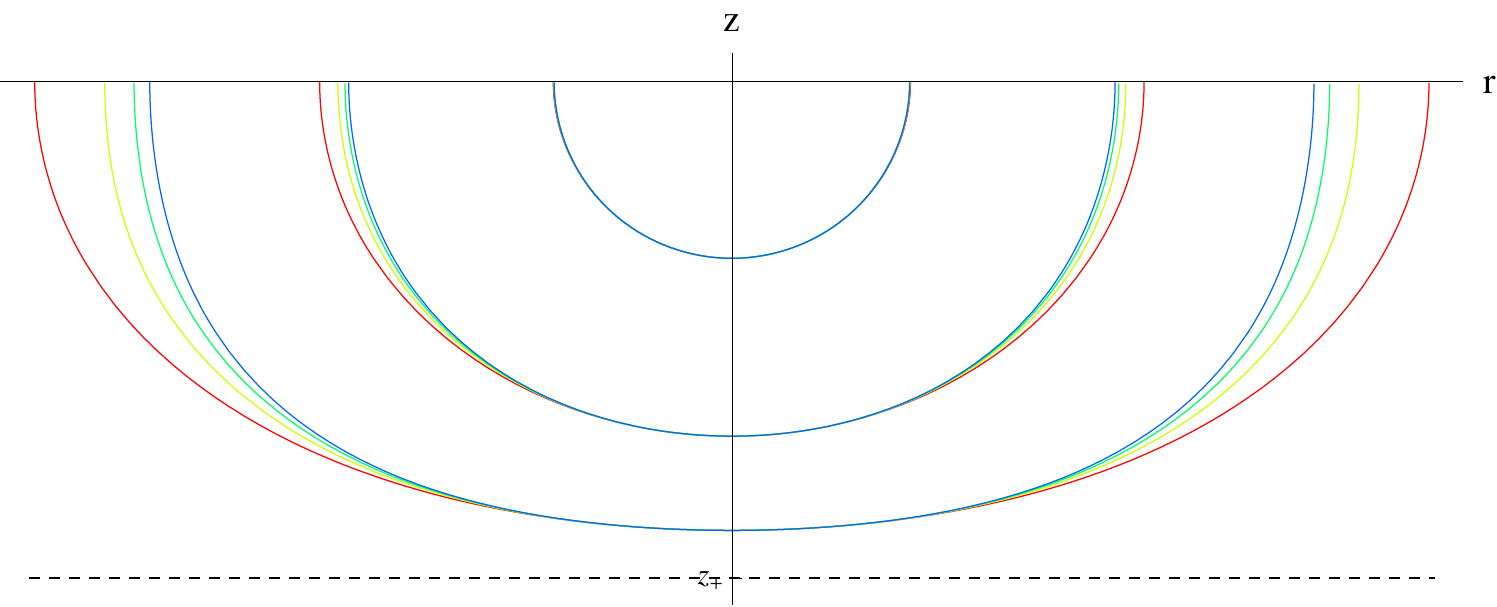}
\caption{
Radial profile of $n$-dimensional extremal surfaces anchored on $n$-ball in a planar Schwarzschild-AdS$_5$ background \req{planarSAdSle}, for various $n$ and various $\zmax$.  The black dashed line at the bottom corresponds to the horizon.  Same parameters and conventions as in \fig{f:minsurfSAdS} are used here.
}
\label{f:minsurfdiskSAdS}
\end{center}
\end{figure}
The resulting surfaces $\CS$ are plotted in \fig{f:minsurfdiskSAdS}, which is the rotationally symmetric analog of the translationally invariant case of \fig{f:minsurfSAdS}.  
The red (outer-most) curves, being geodesics, are of course identical in the two cases, but the higher-dimensional surfaces behave differently.  In particular, in the present case, the profiles of the larger-$n$ surfaces deviate much less significantly from that of the geodesic than was the case for the strip.  This is to be expected, since at any fixed $n$, the `price to pay' for the large area contributions near the boundary is smaller when the perimeter of $\CR$ is smaller.

From  \fig{f:minsurfdiskSAdS} we also see that, not surprisingly, the extremal surfaces anchored on an  $n$-ball on the boundary likewise cannot penetrate the horizon.  In fact, in this case, for fixed extent of $\CR$, while the higher-dimensional surfaces get closer to the horizon than the lower-dimensional ones, the effect is not as pronounced as for the strip case (though the spread between surfaces of various $n$ with fixed $\zmax$ increases as $\zmax \to \zh$).

\subsection{Extremal $n$-surface in Poincare AdS$_{d+1}$ ending on generic regions}
\label{s:ExtGenPoinc}

The motivation for the analysis of the preceding examples was to examine, for given amount of boundary CFT data, which quantities are the best probes of the bulk geometry.
However,  it is not entirely straightforward to compare boundary regions $\CR$ of varying dimensionality.  While the depth $\zmax$  to which their associated extremal surfaces $\CS$ probe has a well-defined meaning in general bulk spacetimes of the form \req{genPoincmetFG}, comparing how much boundary information they require is trickier.  For example, we can't simply compare area of a 2-dimensional region with a volume of a 3-dimensional region. 

One way around this is to define a 1-dimensional quantity for any $n$-dimensional boundary region $\CR$ which characterizes its size.  For example, as mentioned above, we can define the  extent $X(\CR)$ of a region by the maximal distance\footnote{
Recall that when talking about lengths,  volumes, etc., on the boundary, we use the Minkowski metric, i.e.\ we drop the bulk warp factor $\frac{1}{z^2}$.
} between any two points within that region, as indicated in \fig{f:SurfExpl}.  
One can then ask, for a given extent of a region, what region `shape' maximizes the depth $\zmax$ to which the corresponding extremal surface reaches?

Alternately, at fixed $n$, we can compare the various attributes of $n$-dimensional regions of different shapes.  For example, we can ask: for a fixed $n$-volume on the boundary, what shape of  $\CR$ is the most optimal for probing deepest into the bulk?  As we will see momentarily, it turns out that the answers to both of these questions in fact coincide.  For convenience we will then concentrate on the second question, and propose the following conjecture:

\paragraph{Conjecture:}
{\it For a fixed area of a boundary region $\CR$, the shape of $\CR$ which maximizes the reach $\zmax$ of the corresponding extremal surface $\CS(\CR)$ is the round ball.  }

Before providing evidence for the conjecture, let us briefly discuss its implications.  First of all, it is a known geometrical fact that for a given volume of a region, its extent (as defined above) is minimized when the region $\CR$ is the round ball.  Hence if the round ball is the region whose bulk extremal surface $\CS$ reaches deepest into the bulk amongst all regions with fixed volume, then it is also the region whose $\CS$ reaches deepest into the bulk amongst all regions with fixed extent.
In fact, using the results of the preceding subsection, the conjecture then extends to any 
static asymptotically Poincare-AdS bulk geometry with planar symmetry.
In particular, the deepest into the bulk that {\it any} $n$-surface with volume $V_n \, R_0^n$ could reach in {\it any} asymptotically Poincare-AdS bulk of the form \req{genPoincmetFG} is $\zmax = R_0$ where $z$ is the usual Fefferman-Graham coordinate, and $V_n = \pi^{n/2} / \Gamma(\frac{n}{2}+1)$ is the volume of the unit $n$-ball.
In the remainder of this section, we will first give general arguments for our conjecture and then justify it with explicit calculations for $n=2$. 

For the rest of this section, we consider pure Poincare AdS.
The main intuition for the conjecture is built on the following simple observations:
\begin{itemize}
\item Round regions have manifestly greater $\zmax$ than very elongated regions with the same volume.
\item In deforming a ball into e.g.\ an ellipsoid with elongation $\eta$, Poincare AdS does not have enough structure to render $\zmax(\eta)$ a non-monotonic function.  In other words, if a highly elongated ellipsoid yields smaller $\zmax$ than a round ball, then so should one with any elongation.
\item Finally, the dimensionality $n$ of the surface is not crucial for these considerations.
\end{itemize}

The first statement is easy to see by comparing reach of $\CS$ for an $n$-ball with that for a finite strip:  the $n$-ball of radius $R_0$ will have $\zmax = R_0$ (as we found in \sec{s:ExtBallPoinc}) and volume $V=V_n \, R_0^n$.  A `finite strip' of dimensions $(\delta, L \, \ldots, L \} $ will have volume $V = \delta \, L^{n-1}$, so to match the volumes we set $\delta = V_n \, R_0^n / L^{n-1}$.  In the limit  as $L \to \infty$, the ends of the strip become more and more negligible for influencing the shape of the surface away from the ends, so the reach of the surface is better and better approximated by that of an infinite strip of width $\delta$, for which 
\req{nstripzmax} gives  $\zmax = \gamma_n \, R_0^n / L^{n-1} \ll R_0$ where 
$\gamma_n \equiv n \, \pi^{\frac{n-1}{2}} \, \Gamma \left( \frac{2n+1}{2n} \right)
/ (\Gamma \left( \frac{n+1}{2n} \right) \, \Gamma \left( \frac{n+2}{2} \right)) \sim \CO(1)$.

The second expectation implies that we can likewise use surfaces anchored on regions given by small deformations of the round ball.  If by squeezing the ball slightly at fixed volume the corresponding reach $\zmax$ recedes, then we expect that the round ball allows the greatest $\zmax$ among all regions with fixed volume.  This motivates the following explicit calculation: suppose we deform a circular region $\CR_0$ slightly, e.g.\ make it ellipsoidal.  How does the corresponding extremal surface respond?  This question is amenable to perturbation theory: we linearize the Euler-Lagrange equations defining the surface and solve them around the hemisphere to the requisite order.

The last observation then suggests that it suffices to consider 2-dimensional surfaces; so in what follows, we will take $n=2$.  We would like to fix the area of the 2-dimensional region $\CR$ on the boundary, and ask what shape allows for largest bulk reach $\zmax$ of the corresponding extremal surface.  In practice, it will turn out simpler to fix $\zmax$ and see how the area of $\CR$ responds to changing the shape.  However, we can easily convert to the fixed-area set-up using scale invariance: since we are working in pure AdS, rescaling the area by a factor $\alpha$ simply rescales $\zmax$ by $\sqrt{\alpha}$.

\paragraph{EoM for arbitrary 2-surface:}
We can generalize our construction of \sec{s:ExtBallPoinc}, which for surfaces of revolution $z(r)$ gave the Euler-Lagrange equations \req{ELzrPoinc}, to consider surfaces parameterized by $z(r,\phi)$.  
Denoting $\zd \equiv \frac{\partial z}{\partial r}$, $\zp \equiv \frac{\partial z}{\partial \phi}$, etc., we obtain the area of the surface $\CS(\CR)$, analogous to \req{AzrPoinc}, as
\begin{equation}
A = \int_0^{2 \, \pi}  \int_0^{R(\phi)}
	\frac{\sqrt{1+\zd(r,\phi)^2 + \frac{\zp(r,\phi)^2}{r^2}}}{z(r,\phi)^{2}} \, r \, dr \, d\phi
\label{AzrphiPoinc}
\end{equation}	
which yields the following equation of motion:
\begin{equation}
\zdd \left(r^2 + \zp^2 \right) 
+ \zpp \, \left(1+\zd^2 \right)
- 2 \, \zdp \, \zd \, \zp
+\frac{2\, r^2}{z}   \left( 1 + \zd^2 + \frac{\zp^2}{r^2} \right)
+ r \, \zd \,  \left( 1 + \zd^2 + \frac{2 \, \zp^2}{r^2} \right)
= 0 \ .
\label{}
\end{equation}	
It is trivial to check that the hemisphere $z(r,\phi) = \sqrt{\zmax^2 - r^2}$ is indeed a solution.

However, linearization about the hemisphere breaks down near the boundary where we wish to deform $R(\phi)$ away from a constant, since $\zd$ diverges there.
In other words, $\{ z, r, \phi \}$ is not a good set of coordinates for this problem.  

To surmount the difficulty encountered in trying to specify the surface by $z(r,\phi)$, we need to choose a more convenient set of coordinates, namely ones which are better adapted to the  $0^{\rm th}$ order solution.
The most natural option is to use spherical polar coordinates, defined by
\begin{equation}
x = \rho \, \sin \theta \, \cos \phi \ , \qquad
y = \rho \, \sin \theta \, \sin \phi \ , \qquad
z = \rho \, \cos \theta 
\label{}
\end{equation}	
and describe our surface in terms of these by specifying $\rho(\theta,\phi)$.  
In particular, the $0^{\rm th}$ order solution is simply a constant, $\rho(\theta,\phi) = \rho_0$.

Let us first find the full equation of motion for this surface.
Poincare AdS is written as
\begin{equation}
ds^2 = \frac{1}{\rho^2 \, \cos^2 \theta} \, \left[ -dt^2 + d\rho^2 
+ \rho^2 \, \left( d \theta^2 + \sin^2 \theta \, d \phi^2 \right)
+ \sum_{j=3}^{d-1} d\tilde{y}_j^2 \right] \ ,
\label{AdSPoincmetrho}
\end{equation}	
so our Lagrangian (as usual given by the square root of the determinant of the induced metric on the surface) is
\begin{equation}
{\cal L}(\rho,\rhd,\rhp;\theta,\phi) 
= \frac{\sqrt{ \sin^2 \theta \, \left( \rho^2 + \rhd^2 \right) + \rhp^2 }}
	{\rho \, \cos^2 \theta} \ ,
\label{}
\end{equation}	
which gives the  resulting equation of motion 
\begin{equation}
\rhdd \left(\rhp^2 + s^2 \, \rho^2 \right)
+ \rhpp \left(\rhd^2 + \rho^2 \right)
- 2 \, \rhdp \, \rhd \, \rhp
+ s^2 \, \rhd^2  \left( \frac{s^2+1}{s \, c} \rhd - \rho \right) 
+\rhp^2\left( \frac{2}{s \, c} \rhd - \rho \right)
+\rhd \, \rho^2 \, \frac{s}{c}(s^2 + 1)
=0
\label{ELrho}
\end{equation}	
where we've used the shorthand notation $s \equiv \sin \theta$, $c \equiv \cos \theta$, 
 $\rhd \equiv \frac{\partial \rho}{\partial \theta}$, $\rhp \equiv \frac{\partial \rho}{\partial \phi}$, $\rhdp \equiv \frac{\partial^2 \rho}{\partial\theta \, \partial \phi}$, etc..
 
Since all terms in \req{ELrho} come with at least one derivative, it is clear that the hemisphere 
$\rho(\theta,\phi) = \rho_0$ solves this equation.
To find a more general solution, we now linearize and solve \req{ELrho} order by order.

\paragraph{Linearized Solution:}
Let us linearize \req{ELrho} to second\footnote{
It will become clear momentarily why going to just the first order does not suffice.}
 order around the hemisphere solution.
In particular, for
\begin{equation}
\rho(\theta,\phi) = 
\rho_0 + \eps \, \rho_1(\theta,\phi) + \eps^2 \, \rho_2(\theta,\phi) + \CO(\eps^3)
\label{rhoexp2}
\end{equation}	
the $\CO(\eps)$ part of \req{ELrho} gives the equation
\begin{equation}
s^2 \, \rhdd_1 + \rhpp_1 +  \frac{s}{c}(s^2 + 1) \, \rhd_1 = 0 \ ,
\label{ELrhoLin1}
\end{equation}	
while the $\CO(\eps^2)$ part of \req{ELrho} gives the equation
\begin{equation}
s^2 \, \left( \rho_0 \, \rhdd_2 + 2\,  \rho_1 \, \rhdd_1 - \rhd_1^2 \right)
+ \left( \rho_0 \, \rhpp_2 + 2\,  \rho_1 \, \rhpp_1 - \rhp_1^2 \right)
+  \frac{s}{c}(s^2 + 1) \, \left( \rho_0 \, \rhd_2 + 2 \,  \rho_1 \, \rhd_1 \right)
=0 \ .
\label{ELrhoLin2}
\end{equation}	

We can easily solve \req{ELrhoLin1} by separation of variables.  
In fact, we can find a family of solutions, labeled by the mode $\ell \in \ZZ$:
\begin{equation}
\rho_1(\theta,\phi) 
= \tan^\ell(\theta/2) \, (1 +  \ell \cos \theta ) \, \cos \ell \phi \ .
\label{rho1}
\end{equation}	
Let us examine this solution in more detail.
Since $\rho_1(\theta = 0,\phi) = 0$, the maximal reach of the deformed surface remains unmodified, $\rho(\theta = 0,\phi) = \rho_0$.  On the other hand, the boundary region on which this surface is anchored is deformed:  it is now described by 
\begin{equation}
R(\phi) = \rho(\theta = \frac{\pi}{2},\phi)
 = \rho_0 + \eps  \, \cos \ell \phi  \ .
\label{}
\end{equation}	
This means that the area likewise gets modified:
\begin{equation}
A(\CR) = \frac{1}{2} \int_0^{2 \pi} R(\phi)^2 \, d\phi 
= \pi \, \rho_0^2 \, \left(1+ \frac{\eps^2}{2 \rho_0^2} \right)
\label{}
\end{equation}	
(independently of $\ell$).
However, since the area gets modified only at $\CO(\eps^2)$, we need to solve the equation of motion to $\CO(\eps^2)$.

Substituting the first order solution \req{rho1} into the second order equation \req{ELrhoLin2}, we see that although \req{ELrhoLin2} is no longer separable, the only $\phi$-dependent terms are proportional to $\cos 2\ell\phi$, which allows us to write a simple ansatz $\rho_2(\theta,\phi)  = h(\theta) + k(\theta) \, \cos 2\ell\phi$.
At $\phi = \frac{\pi}{4 \ell}$ all $k$-dependence disappears, so we can solve \req{ELrhoLin2} for $h$, fixing constants of integration using regularity.  We can choose another value of $\phi$ (in practice  $\phi = \frac{\pi}{2 \ell}$ takes the most convenient form) to solve for $h$, again  fixing one of the constants of integration using regularity.  The other constant remains undetermined and gives us a second parameter (which we'll denote by $\mu$) to describe the second order solution.
In particular, the second order solution, labeled by integer $\ell$ and real number $\mu$,  is:
\begin{equation}
\rho_2(\theta,\phi) 
= \frac{1}{4 \, \rho_0} \, \tan^{2 \ell}(\theta/2) \, 
 \{ (1 +  \ell \cos \theta )^2 
 + \left[ \mu \, (1 + 2\, \ell \cos \theta ) + \ell^2 \, \cos^2 \theta \right] \, \cos 2 \ell \phi \}
\label{rho2} \ .
\end{equation}	
It can be easily checked that the solution \req{rhoexp2} with the 1st and 2nd order terms given by \req{rho1} and \req{rho2} respectively does solve the equation of motion \req{ELrho} at $\CO(\eps^2)$.
Moreover, we can check that since  $\rho_2(\theta = 0,\phi) = 0$, the maximal reach of the deformed surface still remains unmodified, $\rho(\theta = 0,\phi) = \rho_0$ to $\CO(\eps^2)$.
The corresponding boundary region is now given by 
\begin{equation}
R(\phi) = \rho(\theta = \frac{\pi}{2},\phi)
 = \rho_0 + \eps  \, \cos \ell \phi + \frac{\eps^2}{4\rho_0} (1 + \mu \, \cos 2 \ell \phi) \ .
\label{Rmuell}
\end{equation}	
Computing its area,
\begin{equation}
A(\CR) = \frac{1}{2} \int_0^{2 \pi} R(\phi)^2 \, d\phi 
= \pi \, \rho_0^2 \, \left(1+ \frac{\eps^2}{\rho_0^2}
+ \frac{\eps^4}{\rho_0^4} \frac{(2 + \mu^2)}{32}  \right) \ ,
\label{}
\end{equation}	
we find that $A$ is again independent of $\ell$, and to $\CO(\eps^2)$  also independent of $\mu$.
Since the solution is valid to $\CO(\eps^2)$, we can now trust the area up to $\CO(\eps^2)$ as well.  

Hence we can now unequivocally conclude that for a fixed $\zmax = \rho_0$, the 2-parameter family of deformed hemispheres anchored on the regions $\CR$ given by \req{Rmuell} all have larger area than the round disk.  Conversely, if we keep the area of $\CR$ fixed, the reach $\zmax$ is maximized for the round disk, and is strictly less for any deformation of the disk.  This proves our conjecture at the perturbative level.
\begin{figure}[t!]
\begin{center}
\includegraphics[width=6in]{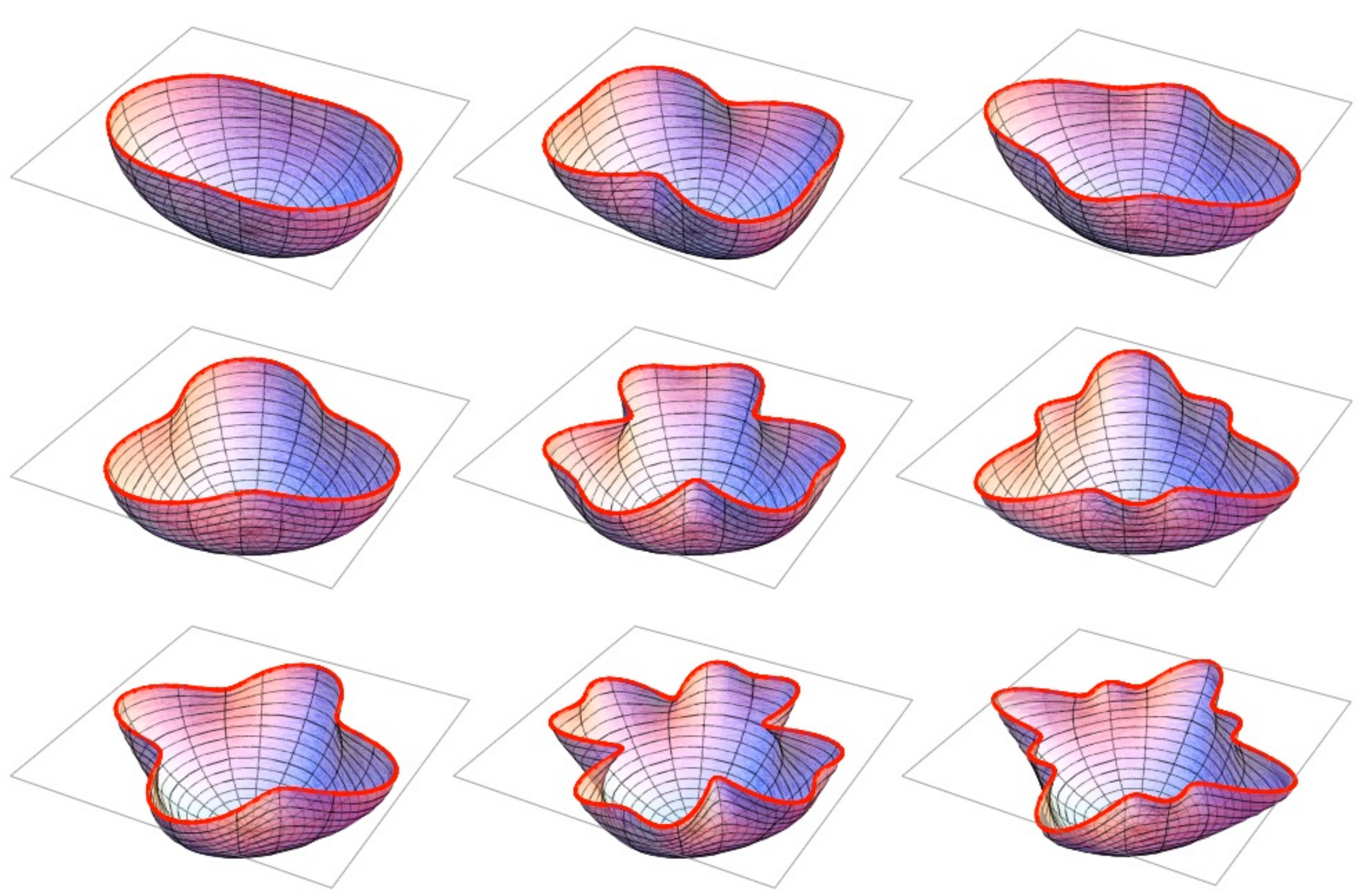}
\caption{
Various extremal surfaces, anchored on a boundary region $\CR$ (bounded by the thick red curve).  In all cases, area of $\CR$ is set to unity and to guide the eye, we show boundary square of with sides $\Delta x = \Delta y =1/\sqrt{2}$  (denoted by the gray quadrilateral).  In all 9 cases, we take 
$\eps/\rho_0 = 1/4$.
The cases are distinguished by values of $\ell$ and $\mu$; in top, middle and bottom rows, $\ell = 2,3,4$, respectively; whereas in left, middle, and right columns, $\mu = 0, -10, 10$, respectively.
}
\label{f:MinSurfVarRegs}
\end{center}
\end{figure}
 \fig{f:MinSurfVarRegs} explicitly shows the various extremal surfaces given by \req{rhoexp2} with \req{rho1} and \req{rho2}, to illustrate the wide variety of extremal surfaces our analysis encompasses. 

One can also check our conjecture at the fully non-linear level.  Exploring various cases numerically\footnote{We thank Padmini Rangamani for explicit check of $\CS$ for $\CR$ having a square and an elongated ellipse shape using the package ComSol.} confirms that the round disk indeed reaches furthest into the bulk.
One should also be able to prove the conjecture mathematically, perhaps using symmetrization techniques.\footnote{We thank Norbert Peyerimhoff for discussions on this issue.}

\section{Extremal surfaces cannot penetrate horizons}
\label{s:ExtSurfHor}

So far, we were mostly interested in comparing various kinds of extremal surfaces as probes of a given spacetime.  The primary motivation in such study was to determine which CFT quantities are most useful in learning about the bulk geometry.
We now turn to perhaps the most interesting aspect of probing the bulk, by asking which of these bulk probes can actually see past an event horizon of the bulk geometry.  It is clear  that no probe corresponding to a causal quantity, such as null geodesics (which determine the bulk-cone singularities of the CFT  \cite{Hubeny:2006yu}), can reach past the horizon, by purely causal considerations.  On the other hand, there is a-priori no such obstruction for spacelike curves or surfaces.  Nevertheless, as we saw in \sec{s:statBH} for geodesics and in \sec{s:ExtStripAsPoinc} and \sec{s:ExtBallAsPoinc} for specific classes of surfaces, these particular spacelike probes do not reach past the horizon.  In this section we set out to show
 that extremal surfaces cannot reach past an event horizon of a static bulk geometry
 in far greater generality, for any dimension $n\le d-1$ of the surface and any shape boundary region $\CR$, as well as for any spacetime with the requisite symmetries.

We will consider static asymptotically AdS spacetimes with planar symmetry described by\footnote{
Since this section is self-contained and we will not need to compare with the Fefferman-Graham form, for notational convenience we now drop the tilde on $\znFG$ in \req{genPoincmet},  and write the general metric simply as  \req{genPoincmet2}.
} \req{genPoincmet} with arbitrary $f$ and $h$,
\begin{equation}
ds^2 = \frac{1}{z^2} \, \left[ -f(z) \, dt^2 +  dx_i \, dx^i 
+ h(z) \, dz^2 \right] \ .
\label{genPoincmet2}
\end{equation}	
Let us for definiteness assume that the geometry \req{genPoincmet2} has a non-degenerate event horizon at $z=\zh$. Then $h$ gets arbitrarily large as we approach the horizon: $f(z\to\zh)\to0$, so $h(z\to\zh^{\mp}) \to \pm \infty$.

Moreover, we will consider arbitrary $n$-dimensional simply-connected region $\CR$ on the boundary.  The corresponding extremal surface $\CS$ can then be described by $z(x^1,\ldots,x^n)$, and its equation of motion is obtained by extremizing its $n$-area, obtained as usual by integrating the square root of the determinant of the induced metric  on $\CS$.
The induced metric $G_{ij}$  is given by
\begin{equation}
G_{ij} = \frac{1}{z^2} \left[ \delta_{ij} + h(z) \, z_{,i} \, z_{,j} \right]
\label{}
\end{equation}	
where we have introduced the notation $z_{,i} \equiv \frac{\p z}{\p x^i}$.
The determinant then takes the simple form
\begin{equation}
G 
	\equiv \varepsilon^{i j \ldots k} \,  G_{1i} \, G_{2j} \ldots G_{nk}
	= \frac{1+ h(z) \left( z_{,1}^2 + \ldots + z_{,n}^2 \right) }{z^{2n}}
\label{Ggensurf}
\end{equation}	
since all the `off-diagonal' contributions cancel by (anti)symmetry.

Already at this level, we can motivate why the extremal surface $\CS$ cannot reach past the horizon, i.e.\ why $\zmax < \zh$.  For if it did, then $G$ would flip sign (since $h$ flips sign and by virtue of $\CS$ `reaching' though the horizon, $z$ must vary, so at least one of the $z_{,i}$'s in \req{Ggensurf} is nonzero).  But the sign of $G$ determines the signature of $\CS$: for a spatial surface, $G>0$.  Since extremal surfaces which are spacelike near the boundary must remain spacelike everywhere, it is clear that $\CS$ cannot reach past $\zh$.  

Since the above argument may seem a bit slick, and is not as obvious in case of degenerate horizons, let us consider the problem at the level of the equation of motion, which will provide further insight into how extremal surfaces always get `repelled' by the horizon.
The lagrangian for the problem is
\begin{equation}
\CL(z \, , \, z_{,1} \, , \, \ldots \, , \, z_{,n} \, ; \, x^1,\ldots,x^n) = \sqrt{G}
	= \frac{\sqrt{1+ h(z) \left( z_{,1}^2 + \ldots + z_{,n}^2 \right) }}{z^{n}}
\label{}
\end{equation}	
and to find the extremal surface, we need to solve the Euler-Lagrange equation,
\begin{equation}
\frac{\p \CL}{\p z} =
  \frac{\p}{\p x^1} \left[ \frac{\p \CL}{\p z_{,1}} \right] + \ldots
  + \frac{\p}{\p x^n} \left[ \frac{\p \CL}{\p z_{,n}} \right] 
\label{}
\end{equation}	
for $z(x^1,\ldots,x^n)$.
Explicitly, this equation of motion takes the  form
\begin{equation}
\sum_i z_{,ii} \, \left( 1+ h(z) \sum_j z_{,j}^2 \right)
	- h(z) \, \sum_{i,j} z_{,ij} \, z_{,i} \, z_{,j}
	+ \sum_i z_{,i}^2 \left( \frac{n}{z} + \frac{h'(z)}{2 \, h(z)} \right)
	+ \frac{n}{z \, h(z)} 
	=0 
\label{ELarbSurf}
\end{equation}	
where $z_{,ij} \equiv \frac{\p^2 z}{\p x^i \, \p x^j}$.

Now let us consider the surface $\CS$ near its deepest point (i.e.\ global maximum of $z$), $z=\zmax$.
At this point, we have $z_{,i}(\zmax)=0$, so evaluating \req{ELarbSurf} at $\zmax$, we find that most of the terms vanish and we are left with
\begin{equation}
\sum_i z_{,ii}+ \frac{n}{z \, h(z)} 
	=0 \qquad {\rm at } \ \ z=\zmax \ .
\label{ELarbSurfzmax}
\end{equation}	
In order for $z=\zmax$ to be a maximum, we additionally need
$z_{,ii}(\zmax)<0$.
This forces $h(\zmax)>0$, i.e.,
$\zmax < \zh$.  In other words, the maximal reach of $\CS$ cannot not extend all the way to the horizon, which shows that extremal surfaces cannot penetrate horizons.

Moreover, the closer $\CS$ approaches to the horizon, the smaller magnitude $z_{,ii}$ may be whilst satisfying \req{ELarbSurfzmax}, so the flatter $\CS$ has to get near its tip -- but that in turn means that the larger the corresponding boundary region $\CR$ has to be.
As the tip of the extremal surface approaches the horizon, the extent of the boundary region $\CR$ spanned by $\p \CS$ diverges.
We now see that this argument in fact generalizes to any horizon, including the degenerate case.

\section{Discussion}
\label{s:Discussion}

In this paper we have explored the capacity  of various CFT probes for encoding the bulk geometry.  The probes under consideration were those represented by specific geometric quantities, namely extremal surfaces, in the bulk.  Since the position of an extremal surface is determined by the bulk geometry and the boundary conditions, one may hope to invert this information to learn about the bulk geometry using the CFT data.  In particular, using a family of boundary conditions and some characteristic of the surface such as its volume, we would hope to extract the bulk metric.  This has been successfully used in several contexts in the past, as reviewed in  \cite{Hubeny:2010ry}.

The goal of the present work was not to perform such an extraction, but rather to explore its limitations. In particular, one may  extract the bulk geometry by extremal surface probes only in a region of spacetime which is accessible to such extremal surfaces.  
Considering which types of surfaces are {\it a-priori} (i.e.\ before knowing the bulk geometry) most suitable for reaching the largest region of the bulk,  then allows us to focus on the most convenient CFT data to be used for our extraction.  Of course, there is a limitation to how far one can get while maintaining full generality.  We have therefore focused on a case where the bulk geometry has certain symmetries which can be easily specified within the CFT.

In particular, if the CFT state is static and homogeneous, the bulk geometry likewise inherits these properties; so in the present work, we have considered static spacetimes which were homogeneous in the boundary directions.  
We have considered two classes of boundary background spacetimes on which the field theory lives: in \sec{s:geods} we considered CFT on Einstein static universe relevant for spherically symmetric  asymptotically globally AdS spacetimes \req{metgen}, while in \sec{s:ExtSurf} and \sec{s:ExtSurfHor} we turned to CFT on Minkowski spacetime, pertaining to translationally invariant  asymptotically Poincare AdS bulk geometries \req{genPoincmet}.

Apart from rendering the calculations tractable, this restriction had the advantage of enabling us to characterize and compare the reach of the various probes in a simple geometrical fashion.  In the former case of global AdS we defined the radial coordinate $r$ by proper area of the $d-1$ spheres as usual, and characterized the reach of extremal surfaces -- in that case geodesics -- by the minimum value $\rmin$ that this coordinate achieves along a given geodesic. In the latter case, we used Fefferman-Graham coordinate $z$ (related to proper length along radial geodesics) and characterized the reach of extremal surfaces by its maximal value $\zmax$ attained along such surfaces.

Although one could relax our symmetry constraints it would complicate the comparisons. For example, in a spatially-varying geometry, one semi-natural way to define $\zmax$ purely geometrically would be to use affine parameter along ingoing geodesics emanating orthogonally to the boundary fluid velocity, analogously to the set-up used in the fluid/gravity correspondence \cite{Bhattacharyya:2008jc}; however unless the spacetime varied slowly enough in the boundary directions, the caustics where such coordinate system breaks down might occur already within the reach of the extremal surfaces.
While by no means insurmountable, this complication would nevertheless hinder us in making  fully general statements.  

On the other hand, it would be very interesting to explore the consequences of relaxing the imposed symmetries.  For instance, one could explore whether the extremal surfaces deform in some intriguing fashion.  Of particular interest would be to introduce some non-trivial time dependence into our spacetime.  As mentioned below, this can even invalidate the conclusions reached in the previous section by allowing extremal surfaces to extend through an event horizon.  One convenient set-up wherein one could make definitive statements in absence of any symmetries is the fluid/gravity regime
\cite{Bhattacharyya:2008jc}, valid when the geometry varies sufficiently slowly in the boundary directions.  This is unfortunately unlikely to produce the type of evolution needed to probe past event horizons, but nevertheless would allow us to gain further insight into the effects that time dependence can produce. 

A much milder form of relaxing staticity was  explored in \cite{Hubeny:2007xt} by considering stationary geometries.  While such geometries are still time-translation invariant, they are not time-reversal invariant, so extremal surfaces need no longer lie at constant-$t$ slice.  For example,  the set of all spacelike geodesics anchored at $t=0$ on the boundary of  rotating BTZ spacetime spans not a single spacelike slice of the bulk but rather a co-dimension 0 spacetime volume with finite time extent.  This suggests the intriguing possibility that the knowledge of the CFT data at a single time allows us to directly extract the bulk geometry over a range of bulk times.
These observations demonstrate that relaxing the symmetries allows a richer set of possibilities, which we leave for future explorations.

Returning to the static and homogeneous configurations discussed above, specified by two arbitrary functions of one variable (the bulk radial coordinate), we now summarize our main results.

In \sec{s:geods} we explored `probe' geodesics, characterized by both endpoints anchored on the (same) AdS boundary.  While null geodesics might represent the easiest CFT probes to use in terms of ready availability of the CFT data, we saw that they are not the best probes for accessing the largest possible region of spacetime.  In particular, for any null geodesic, there always exists a spacelike one which admits smaller $\rmin$.  This has implications for decoding the spacetime
if we restrict the CFT region on which we're allowed to specify the CFT data, or
 if the bulk spacetime is causally nontrivial (or  admits null circular orbits).
In such cases, the `optimal' geodesics to probe the spatial geometry turn out to be the $E=0$ spacelike ones\footnote{
These are however  insensitive to the time component of the metric as they lie at constant $t$.
} since these not only have $\Delta t = 0$, but also minimize $\rmin$ at fixed $\Delta \ph$.  

In the causally interesting case describing a black hole, we showed that probe geodesics cannot reach past the horizon, but that there is a family of spacelike geodesics which can probe {\it arbitrarily near the horizon}; these correspond to $E=0$ and $L \to (\rh)^+$.   Such geodesics have large $\Delta \ph$ and correspondingly large $\CL_{\rm reg}$, because they orbit the black hole in the vicinity of the horizon many times.  Imposing finite upper bound on $\Delta \ph$ or $\CL_{\rm reg}$ restricts $\rmin$ to be finitely larger than $\rh$, though we saw that for typical examples this is not too severe.  Null geodesics, on the other hand, can typically probe to only $\CO(\rh)$ distance from the horizon.

Given that $E=0$ geodesics 
represented the best 1-dimensional probes, in the remaining sections we restricted attention to extremal surfaces anchored on the boundary at constant time.  For static geometries this implies that the entire surface lies at constant bulk time, 
and is therefore  insensitive to $g_{tt}$ in the bulk metric. 
This leaves one arbitrary function of the bulk radial coordinate,  characterizing the spatial geometry of the bulk, which we can try to extract via such extremal surfaces.  \sec{s:ExtSurf} explored which extremal surfaces are the best probes of this spatial geometry.  Unlike the geodesic case where we only had a single parameter, $L$, at our disposal, we now have a far richer set of attributes by which to characterize an extremal surface $\CS$.  Apart from the dimensionality $n$, we can freely specify the boundary region $\CR$ defining the surface $\CS$.  In  \sec{s:ExtSurf} we examined how these attributes affect the behaviour of $\CS$.

One theme which we explored was how does the dimensionality $n$ of the extremal surface affect its reach $\zmax$.  
In \sec{s:ExtStripPoinc} we considered  `strip' regions $\CR$ with a given thickness $\Delta x$,   observing that  higher-dimensional surfaces appear to be better probes of the bulk geometry, in the sense that $\zmax/\Delta x$ grows with $n$ in pure AdS.  To convince ourselves  that this is not an artifact of pure AdS geometry, in \sec{s:ExtStripAsPoinc} we extended the result to more general asymptotically AdS spacetimes.
However, as we saw, this question needs to be specified more carefully, since we can't directly compare quantities of different dimensions.  In particular, the strip has infinite extent in the remaining $n-1$ directions, so that higher-dimensional regions utilize much more CFT data.  Considering round $n$-ball regions in \sec{s:ExtBallPoinc} provided a more indicative comparison by specifying the (1-dimensional) extent of $\CR$.  In pure AdS, it turns out that $n$-dimensional extremal surfaces are hemispherical and have the same reach $\zmax$ for all $n$.  In more general asymptotically AdS geometries, as discussed in \sec{s:ExtBallAsPoinc}, it is however still true that higher-dimensional surfaces with fixed extent probe slightly deeper.\footnote{
While we considered static geometries, this observation would hold in dynamically evolving spacetimes which are static outside some region, such as dynamical shell collapse often used as a toy model of thermalization in the CFT.
This makes the recent observation \cite{Balasubramanian:2011ur} (also noted by \cite{Keranen:2011xs} in context of asymptotically-Lifshitz Vaidya spacetimes) that thermalization is slower for entanglement entropy than for two-point functions 
evident from the bulk perspective, for any situation where the non-thermal region implodes.
}
The general lesson then remains, that utilizing more CFT data recovers more information about the bulk geometry.

These results motivated  us to consider the effect that the shape of $\CR$ has on $\CS$.  In \sec{s:ExtGenPoinc}, we saw that the round regions ($\CR$ being the $n$-ball) are in fact `optimal'.  
In particular, we have argued that for a fixed area of a boundary region $\CR$, the shape of $\CR$ which maximizes the reach $\zmax$ of the corresponding extremal surface $\CS(\CR)$ is the round ball.
In this sense,  extremal $n$-surfaces anchored on round $n-1$-spheres on the boundary are the closest higher-dimensional analogs of the $E=0$ probe geodesics of \sec{s:geods}.
The more elongated $\CR$ becomes  at fixed volume, the smaller its reach $\zmax$.

We also explored the effect that spacetime deformations have on extremal surfaces anchored on a given region $\CR$.  We suggested that for `physically sensible' deformations, the reach $\zmax$ is smaller than in pure AdS.  Said more physically, the deeper the gravitational potential well, the less deep into the bulk do the extremal surfaces reach.  We motivated this statement from two directions:  firstly by considering  the leading asymptotic fall-off of the bulk deformation, which is related to the boundary stress tensor, and secondly by noting that
in the extreme case of spacetime with a black hole, the horizon in fact expels the extremal surfaces entirely.
It would be nice to interpolate these two pictures in some sense, and in particular to pinpoint the conjectured physical pathology in the CFT associated with bulk deformation wherein some extremal surfaces reach deeper than the corresponding ones in pure AdS.

Combining the results of the previous three paragraphs, we conclude that in any physically sensible spacetime of the form \req{genPoincmet}, the deepest that an extremal surface $\CS$ of extent $x$ can reach is bounded by $\zmax \le x/2$.  Similarly,  the deepest that an extremal surface $\CS$ of $n$-volume $R_0^n \, \pi^{n/2} / \Gamma(\frac{n}{2}+1)$ can reach is bounded by $\zmax \le R_0$.   These bounds are saturated only for spherical regions in pure AdS: deforming the shape of the boundary region $\CR$ or the spacetime in which the surface $\CS$ lives results in smaller reach.  As we saw in the various examples studied, large deformations decrease $\zmax$ substantially.

The above observations lead us to perhaps the most intriguing result of this paper, namely that extremal surfaces in static spacetimes cannot penetrate event horizons.  We have seen this statement in several different guises and proved it in various contexts using a variety of arguments.
  In \sec{s:statBH}, we demonstrated it in context of probe geodesics by using constants of motion, in \sec{s:ExtStripAsPoinc} we showed it for extremal surfaces anchored on a strip  by using properties of the effective potential, and finally in \sec{s:ExtSurfHor} we proved it more generally by analyzing the equations of motion directly.  The latter allowed us to make the argument for any dimension $n\le d-1$ of the surface and any shape of the boundary region $\CR$, as well as for arbitrary spacetime with the requisite symmetries.

One motivation for deriving this fact from different viewpoints was to get a feel for its robustness.  In particular, we know that it does not hold in fully general spacetimes.
Our arguments crucially relied on the fact that the spacetimes under consideration were static.  In all such cases the event horizon coincides with the Killing horizon and apparent horizon, and its position can be determined locally.  
On the other hand, once we allow the spacetime to evolve in time, these arguments cease to hold.
In fact, it is easy to obtain examples where extremal surfaces {\it do} probe past event horizon, by utilizing the teleological nature of the event horizon.  

One such example was provided early on in \cite{Hubeny:2002dg} by using the following gedanken experiment:  Suppose we collapse a high-energy shell in AdS, in such a way that the shell implodes from the boundary at time $t=0$, collapses, and forms a large black hole with horizon radius $\rh \gg 1$ (in AdS units).  The bulk geometry is pure AdS to the past of the shell and Schwarzschild-AdS to its future, so the event horizon  is generated by  the outgoing radial null geodesics which cross the shell at $r=\rh$.  This means that the event horizon forms at the origin $r=0$ at some earlier time $t_h=-\tan^{-1} \rh \to -\pi/2$ as $\rh \to \infty$.
Now consider a given spacelike probe, such as a spacelike geodesic with $\Delta \ph = \pi$ stretching across AdS at time $t \in (t_h,0)$.  Such a geodesic passes through only the pure AdS part of the geometry, and so remains insensitive to the event horizon.  However, it traverses through the black hole, and therefore probes bulk regions which are causally disconnected from the entire boundary.
Nevertheless, it is also clear in this example that there still exist bulk regions which remain inaccessible to all probe extremal surfaces.

In such time-dependent situations where the event horizon behaves teleologically, it is tempting to ask whether the bulk regions which are inaccessible to probe geodesics and higher-dimensional extremal surfaces can nevertheless  be defined in some nice geometrical way.  One possible candidate for the excluded region might be a suitable quasi-local horizon.  However, natural as this construct may appear, is does not seem to be viable, since quasi-local horizons are defined in a  foliation-dependent manner,\footnote{
For a nice review of quasi-local horizons, see e.g.\ \cite{Booth:2005qc}.  Shortcomings of foliation dependence in the AdS/CFT context were explained e.g.\ in \cite{Figueras:2009iu}.}
whereas bulk probes are oblivious to this information.
(Also in specific Vaidya-type collapse situations, it has been seen explicitly \cite{AbajoArrastia:2010yt} that geodesics can probe past the apparent horizon while being anchored on the boundary.)
It would be interesting to explore this issue further.

\subsection*{Acknowledgements}
\label{acks}

It is a pleasure to thank 
Roberto Emparan,
Gary Horowitz,
Don Marolf, 
Joe Polchinski, and
Mukund Rangamani
for many stimulating discussions during various stages of this project.
I also thank Hong Liu, Mukund Rangamani, and Tadashi Takayanagi for collaborations which inspired the present exploration.
This work was supported in part by an STFC Consolidated Grant ST/J000426/1.

\providecommand{\href}[2]{#2}\begingroup\raggedright\endgroup

\end{document}